\documentclass[twocolumn]{aastex7}
\usepackage{fontawesome}
\usepackage{color}
\usepackage{graphicx}
\usepackage{amsmath, amssymb, bm}
\usepackage{CJK}
\usepackage[version=4]{mhchem} 
\usepackage{multirow}

\shorttitle{Magma-atmosphere interactions on TOI-270~d}
\shortauthors{Nixon et al.}

\submitjournal{the Astrophysical Journal June 30, 2025, accepted October 27, 2025}

\begin{document}

\title{Magma ocean interactions can explain JWST observations of the sub-Neptune TOI-270~d}



\correspondingauthor{Matthew C.\ Nixon}
\email{matthewnixon@asu.edu}

\author[0000-0001-8236-5553]{Matthew C.\ Nixon}
\altaffiliation{51 Pegasi b Fellow}
\affiliation{School of Earth and Space Exploration, Arizona State University, Tempe, AZ, USA}
\affiliation{Department of Astronomy, University of Maryland, College Park, MD, USA}
\email{mcnixon@umd.edu}

\author[0009-0008-1115-8774]{R.\ Sander Somers}
\affiliation{Department of Astronomy, University of Maryland, College Park, MD, USA}
\email{}

\author[0000-0002-2454-768X]{Arjun B.\ Savel}
\affiliation{Department of Astronomy, University of Maryland, College Park, MD, USA}
\email{}

\author[0000-0003-2775-653X]{Jegug Ih}
\affiliation{Space Telescope Science Institute, 3700 San Martin Drive, Baltimore, MD, USA}
\email{}

\author[0000-0002-1337-9051]{Eliza M.-R.\ Kempton}
\affiliation{Department of Astronomy, University of Maryland, College Park, MD, USA}
\affiliation{Department of Astronomy \& Astrophysics, University of Chicago, Chicago, IL, USA}
\email{}

\author[0000-0002-1299-0801]{Edward D.\ Young}
\affiliation{Department of Earth, Planetary, and Space Sciences University of California, Los Angeles, CA, USA}
\email{}

\author[0000-0002-0298-8089]{Hilke E.\ Schlichting}
\affiliation{Department of Earth, Planetary, and Space Sciences University of California, Los Angeles, CA, USA}
\email{}

\author[0000-0002-3286-7683]{Tim Lichtenberg}
\affiliation{Kapteyn Astronomical Institute, University of Groningen, Groningen, The Netherlands}
\email{}

\author[0000-0003-0156-4564]{Luis Welbanks}
\affiliation{School of Earth and Space Exploration, Arizona State University, Tempe, AZ, USA}
\email{}

\author[0000-0001-6315-7118]{William Misener}
\affiliation{Earth and Planets Laboratory, Carnegie Institution for Science, Washington, DC, USA}
\email{}

\author[0000-0002-4487-5533]{Anjali A.\ A.\ Piette}
\affiliation{School of Physics and Astronomy, University of Birmingham, Edgbaston, Birmingham, UK}
\email{}

\author[0000-0002-0413-3308]{Nicholas F.\ Wogan}
\affiliation{Space Science Division, NASA Ames Research Center, Moffett Field, CA, USA}
\email{}

\begin{abstract}
Sub-Neptunes with substantial atmospheres may possess magma oceans in contact with the overlying gas, with chemical interactions between the atmosphere and magma playing an important role in shaping atmospheric composition. Early JWST observations have found high abundances of carbon- and oxygen-bearing molecules in a number of sub-Neptune atmospheres, which may result from processes including accretion of icy material at formation or magma-atmosphere interactions. Previous work examining the effects of magma-atmosphere interactions on sub-Neptunes has mostly been limited to studying conditions at the atmosphere-mantle boundary, without considering implications for the upper atmosphere which is probed by spectroscopic observations. In this work, we present a modeling architecture to determine observable signatures of magma-atmosphere interactions. We combine an equilibrium chemistry code which models reactions between the core, mantle and atmosphere with a radiative-convective model that determines the composition and structure of the observable upper atmosphere. We examine how different conditions at the atmosphere-mantle boundary and different core and mantle compositions impact the upper atmospheric composition. We compare our models to JWST NIRISS+NIRSpec observations of the sub-Neptune TOI-270~d, finding that our models can provide a good fit to the observed transmission spectrum with little fine-tuning. This suggests that magma-atmosphere interactions may be sufficient to explain high abundances of molecules such as H$_2$O, CH$_4$ and CO$_2$ in sub-Neptune atmospheres, without additional accretion of icy material from the protoplanetary disk. Although other processes could lead to similar compositions, our work highlights the need to consider magma-atmosphere interactions when interpreting the observed atmospheric composition of a sub-Neptune.
\end{abstract}


\section{Introduction}\label{sec:intro}

Determining the characteristics of exoplanets whose radii lie between those of the Earth and Neptune (1--4$R_{\oplus}$) is one of the major outstanding challenges in the study of exoplanets today. The very existence of these planets is intriguing, since no analogues for such objects exist in our solar system. Statistical trends determined from demographic studies of this population have provided some initial insight into their possible compositions. Planets with radii less than 4$R_{\oplus}$ orbiting FGK stars have a bimodal radius distribution \citep{Fulton2017}, suggestive of two sub-populations, often labeled ``super-Earths'' ($R_p \lesssim 1.8R_{\oplus}$) and ``sub-Neptunes'' ($R_p \gtrsim 1.8R_{\oplus}$). It has been suggested that super-Earths are typically rocky bodies with little to no atmosphere, while sub-Neptunes are planets with large H/He atmospheres comprising up to a few per cent of their total mass \citep[e.g.,][]{Chen2016}. However, the range of bulk compositions that can explain the masses and radii of sub-Neptunes is far from unique; for example, a significant number of sub-Neptunes orbiting M dwarf stars appear to be consistent with a water-rich, as well as a hydrogen-rich, atmosphere \citep{Luque2022,Rogers2023}.

JWST is transforming our ability to understand the composition of sub-Neptunes by directly probing their atmospheres. Although only a small number of such planets have been observed to date, a wide compositional diversity has been revealed. A number of sub-Neptune observations have revealed high-metallicity ($>100\times$ solar) or water-dominated atmospheres, such as GJ~1214~b \citep{Kempton2023_gj,Gao2023,Nixon2024_gj,Schlawin2024}, GJ~3470~b \citep{Beatty2024} and GJ~9827~d \citep{Piaulet2024}. However, other sub-Neptune atmospheres appear to be H/He-dominated, and consistent with solar metallicity \citep[e.g., TOI-421~b,][]{Davenport2025}. This diversity prompts further investigation into the physical and chemical processes that shape the atmospheres of these objects. Note that, in this work, we use the term ``metallicity'' to refer to the abundance of elements heavier than H/He in the (observable) upper atmosphere.

The sub-Neptune TOI-270~d \citep{Gunther2019} provides a particularly informative case study thanks to its early observationsal campaigns with JWST (GO \#3557, PI Madhusudhan; GO \#3818, PI Gapp; GO \#4098, PI Benneke). The NIRISS/SOSS and NIRSpec/G395H transmission spectra of the planet \citep{Benneke2024,Holmberg2024} enabled detections of CH$_4$, CO$_2$ and H$_2$O in its atmosphere, with a metallicity of 225$^{+98}_{-86} \times$ solar derived by \citet{Benneke2024} from an analysis of the entire spectrum. This metal-rich composition could be a result of the planet forming further out in the disk, beyond the ice line, and thus accreting more icy material before moving inwards to its present orbital location \citep{Mordasini2009}, or of late-stage pollution by icy planetesimals \citep{Lichtenberg2022}. However, it is also possible that evolutionary processes could enrich the metallicity of an initially H/He-dominated atmosphere, without invoking an ice-rich interior. Such processes include photoevaporation \citep{Owen2017,Heng2025}, core-powered mass loss \citep{Gupta2019,Cherubim2025}, and interaction between the atmosphere and a magma ocean \citep{Kite2020,Lichtenberg2021,Schlichting2022}, the latter being the focus of this work.

Sub-Neptunes with rocky interiors and terrestrial planets are expected to possess magma oceans (molten silicate layers) in contact with their atmosphere for a substantial portion of their lifetime \citep{Schaefer2016,Kite2019,Nicholls2024,Tang2025}. Chemical interactions between the molten core and mantle and the gaseous atmosphere can significantly alter the composition of all components \citep{Lichtenberg2025}. For sub-Neptunes, this process is expected to lead to oxidation of the atmosphere and reduction of the core and mantle \citep{Schlichting2022}, as well as a decrease in the atmospheric carbon-to-oxygen ratio (C/O) \citep{Seo2024}. The studies mentioned here focused on determining the composition at the base of the atmosphere, rather than the upper atmosphere that may be probed by spectroscopic observations. Additional processes, such as thermochemical equilibrium at pressures lower than that of the atmosphere-mantle boundary, as well as condensation, vertical mixing and photochemistry, must be considered if we are to test the observable consequences of interactions taking place in planetary interiors.

Motivated by the arrival of JWST observations of sub-Neptunes, efforts are now underway to connect the composition at the base of the atmosphere derived from magma ocean models to the observable upper atmosphere. To date, this work has primarily focused on explaining observations of the sub-Neptune K2-18~b, which has been hypothesized to host a liquid water ocean beneath its atmosphere \citep[e.g.,][]{Madhu2020,Nixon2021,Madhusudhan2023}. \citet{Shorttle2024} found that invoking a magma ocean could explain the apparent depletion of nitrogen in the planet's atmosphere, which had previously been used as a line of evidence for a liquid water ocean \citep{Yu2021,Hu2021,Tsai2021}. However, \citet{Rigby2024} found that a magma ocean was not consistent with observations of K2-18~b, arguing that the magma ocean models are in tension with the bulk parameters of the planet. Regardless of the final interpretation, these works demonstrate the possibility of connecting equilibrium chemistry models of the core, mantle and base of the atmosphere to the upper atmospheric composition, thus enabling the use of atmospheric compositional constraints from spectroscopy to better understand interior processes in sub-Neptunes. For TOI-270~d, although the possibility of a magma ocean has been suggested \citep[e.g.,][]{Benneke2024,Glein2025}, models predicting the atmospheric composition resulting from magma ocean interactions are yet to be constructed.

In this paper, we develop a new framework for connecting predictions of the composition at the base of the atmosphere from the model presented in \citet{Schlichting2022} to the expected upper atmospheric composition. In particular, we are interested in determining whether magma ocean interactions are sufficient to explain the observed upper atmospheric properties of TOI-270~d. In Section \ref{sec:methods}, we describe our methodology in detail, including our choice of modeling approaches for upper atmospheric processes. We present resulting temperature and volume mixing ratio profiles, as well as a direct comparison to observations of TOI-270~d, in Section \ref{sec:results}. Finally, we discuss caveats and possible future developments in Section \ref{sec:discussion}.

\section{Methods}\label{sec:methods}

\begin{figure*}
    \centering
    \includegraphics[width=\linewidth]{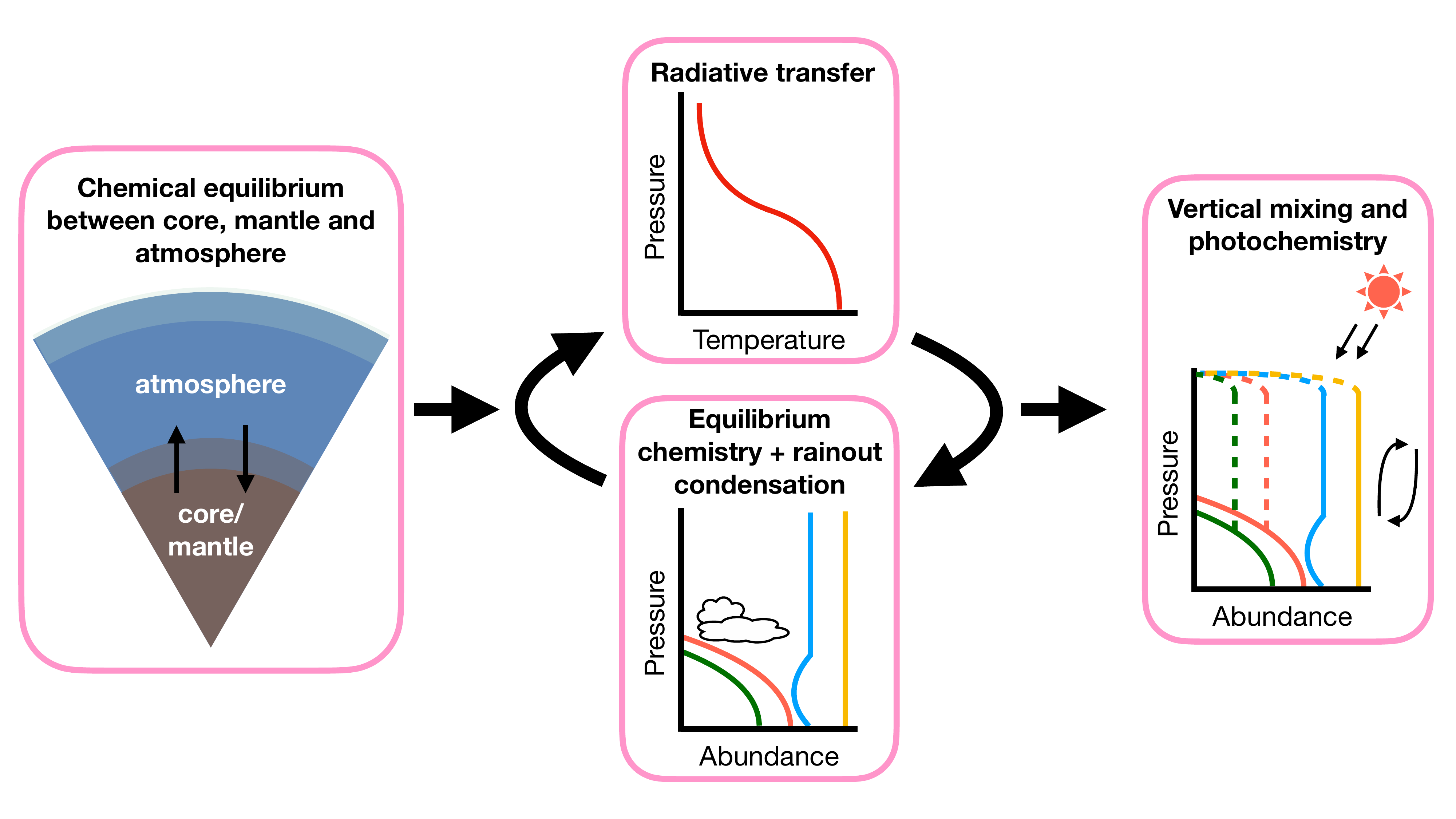}
    \caption{Schematic of the modeling framework used in this study. Chemical equilibrium between the core, mantle and atmosphere is calculated using the model from \cite{Schlichting2022}. Upper atmospheric temperature profiles are calculated using \texttt{HELIOS} \citep{Malik2017}. \textsc{Fastchem cond} \citep{Stock2022} is used to find equilibrium chemical abundances throughout the upper atmosphere, accounting for rainout condensation. After iterating between \texttt{HELIOS} and \textsc{Fastchem cond} until convergence is reached, we account for photochemical processes and vertical mixing using \textit{Photochem} \citep{Wogan2023,Wogan2025}.}
    \label{fig:schematic}
\end{figure*}

Our modeling framework is shown in Figure \ref{fig:schematic}. We use the model described in \cite{Schlichting2022} to calculate chemical equilibrium between the core, mantle and atmosphere. This informs our inputs to \texttt{HELIOS} \citep{Malik2017}, which we use to simulate atmospheric temperature profiles. We iterate between \texttt{HELIOS} and \textsc{fastchem cond} \citep{Stock2022}, which solves for equilibrium chemistry including rainout condensation, until converged temperature and chemical abundance profiles are obtained. We subsequently model photochemistry and vertical mixing using \textit{Photochem} \citep{Wogan2023} before generating transmission spectra with \textsc{Aura-3D} \citep{Nixon2022}. We describe each step in more detail below.

\subsection{Magma ocean-atmosphere chemical interactions}

\citet{Schlichting2022} presented an equilibrium thermodynamic model for sub-Neptunes that accounts for chemical interactions between an iron-rich core, silicate mantle, and hydrogen-rich atmosphere. We use this model to determine the composition at the atmosphere-mantle boundary, which informs our upper atmospheric model. This model uses a variety of sources of thermodynamic data in order to calculate standard-state molar Gibbs free energies of reaction, including: \citet{Fegley1987,Pan1991,Okuchi1997,Moore1998,Hirschmann2012,Badro2015,Hirschmann2016}. Details on which sources were used for each chemical reaction in the model may be found in the Appendix of \citet{Schlichting2022}.

We adopt planetary parameters similar to those of TOI-270~d, whose total iron and silicate mass fraction was determined to be about 90\% (4.3$\pm$0.5$\,M_{\oplus}$) of its total mass, 4.78$\pm$0.43$\,M_{\oplus}$ \citep{VanEylen2021,Benneke2024}. We therefore adopt a total iron and silicate mass of 4.3$\,M_{\oplus}$. We also test the sensitivity of our results to the exact iron and silicate mass (see Section \ref{subsec:abundances}). Following \citet{Schlichting2022}, we consider a reactive core consisting of pure Fe beneath a silicate mantle consisting of 92.1\% MgSiO$_3$, 3.5\% FeSiO$_3$ and 3.2\% MgO by mole fraction, with trace amounts of Na$_2$O, SiO$_2$, Na$_2$SiO$_3$ and FeO. These components react with an atmosphere consisting of approximately solar composition \citep{Asplund2009} with the following volume mixing ratios: $\sim$99.9\% H$_2$, 0.05\% CO and H$_2$O, 10$^{-7}$ CO$_2$ and CH$_4$, 10$^{-9}$ O$_2$, Fe, Mg, Na, SiO and SiH$_4$. Since He is not included in the model from \citet{Schlichting2022}, and is generally unreactive, we do not include it at this stage of the modeling, but introduce He at solar abundance in our radiative-convective models of the upper atmosphere.

We construct a grid of models by varying three different input parameters: $T_{\rm m-a}$, the equilibration temperature at the mantle-atmosphere boundary; $T_{\rm c-m}$, the equilibration temperature at the core-mantle boundary; and $x_{\rm Fe}$, the iron mass fraction of the (molten) iron/silicate nucleus of the planet ($x_{\rm Fe}=m_{\rm Fe}/(m_{\rm Fe}+m_{\rm silicates})$). We consider three combinations of temperatures at the mantle-atmosphere and core-mantle boundaries: (1) $T_{\rm m-a}=2000\,$K, $T_{\rm c-m}=3000\,$K, (2) $T_{\rm m-a}=2000\,$K, $T_{\rm c-m}=4000\,$K, and (3) $T_{\rm m-a}=3000\,$K, $T_{\rm c-m}=4000\,$K (see Table \ref{tab:params}). The choice of temperatures was motivated by extending an adiabatic temperature profile from the base (1000~bar) of a \texttt{HELIOS} model with parameters consistent with observations of TOI-270~d \citep{Benneke2024} to likely pressures at the base of the atmosphere \citep[$\sim6\times10^4$~bar,][]{Schlichting2022}, accounting for the wide uncertainty in the temperature gradient and location of the mantle-atmosphere and core-mantle boundaries. For this initial model, we assumed an atmospheric metallicity of 200$\times$~solar, with a solar C/O, and all other parameters equal to those described in Section \ref{subsec:rc}. We note that temperatures much higher than 5000~K are possible at the core-mantle boundary \citep{Ginzburg2016}; however, calculations at such high temperatures would require unreasonable extrapolation of experimental data \citep{Schlichting2022}. We also consider several values of $x_{\rm Fe}$: 50\%, 33\%, 20\% and 1\%.

\subsection{Upper atmospheric chemistry} \label{subsec:fastchem}

The output of the global chemical equilibrium code described above includes the volume mixing ratios for a range of chemical species in the atmosphere after interactions with the core and mantle. This represents the composition at the base of the atmosphere, which for sub-Neptunes is likely to be located at pressures $\gtrsim$1000 bar, and possibly as high as $\gtrsim$10$^6$ bar \citep{Breza2025}. However, it is unlikely that this is the same as the composition throughout the upper, observable region of the atmosphere ($P\lesssim$1 bar) for several reasons described below. The lower pressures and temperatures of the atmosphere will lead to different chemical species being favored in thermo-chemical equilibrium for a given set of elemental abundances. Furthermore, several of the gas-phase species present in the hot, high-pressure region of the atmosphere \citep[e.g., SiH$_4$,][]{Misener2023} may condense into solid or liquid phases before reaching the observable atmosphere. Finally, photochemistry and vertical mixing will alter the atmospheric composition near the top of the atmosphere away from thermo-chemical equilibrium expectations. It is therefore necessary to connect the global equilibrium chemistry model to a model of the upper atmosphere in order to determine its observable atmospheric properties.

In order to determine the gas-phase chemistry throughout the atmosphere, we first break up the chemical species at the base of the atmosphere (i.e., the outputs of the \citeauthor{Schlichting2022} \citeyear{Schlichting2022} chemical equilibrium model) into constituent elements. The elemental abundances are then used as an input to \textsc{fastchem cond} \citep{Stock2018,Stock2022,Kitzmann2024}, a chemical equilibrium code that includes both equilibrium and rainout condensation. Equilibrium condensation treats condensation at each temperature and pressure point independently. However, this is not well-suited to computing the composition of planetary atmospheres, since rainout will impact the distribution of elements available in the atmosphere as a function of altitude \citep{Burrows1999}. 

In order to simulate the effect of rainout, \textsc{fastchem cond} starts by calculating the chemistry at the highest pressure in the atmosphere and works its way towards lower pressures throughout a given pressure-temperature profile. Whenever condensation is encountered, the coupled condensation and gas-phase system is solved, yielding the effective elemental abundances of the condensing elements left in the gas phase \citep[see equation 12,][]{Kitzmann2024}. These effective abundances are subsequently used to change the actual elemental abundances to the newly-computed values at all pressures below the level at which condensation occurs. This impacts the abundances of elements remaining in the gas phase in the upper atmosphere, and can lead to a sudden decrease in the abundances of condensing elements. Calculation of the condensed phase requires temperature-dependent equilibrium constants, for which \textsc{fastchem cond} uses a variety of data sources: \citet{Prydz1972,Haar1978,Goodwin1985,Sharp1990,Moses1992,Barin1995,Chase1998,Yaws1999,Dykyj2001,Murphy2005,Wagner2008,Lide2009,Gail2013}. Details regarding which sources are used for a given species can be found in Table A1 of \citet{Kitzmann2024}.

We note that the \citet{Schlichting2022} model only includes the following elements: hydrogen, carbon, oxygen, iron, magnesium, silicon and sodium. In general, we set the abundances of any other elements included in \textsc{fastchem cond} to solar composition values as described in \citet{Asplund2009}, thereby assuming that the abundances of these elements are not affected by magma ocean interactions, a simplification that is necessary in order to couple the models. Other works have noted that nitrogen, which is not included in our magma-atmosphere interaction model, may be depleted  relative to chemical equilibrium in the presence of a magma ocean \citep{Shorttle2024}. In order to test whether the lack of nitrogen chemistry in the \citet{Schlichting2022} impacts our results, we perform a sensitivity test by running models with a range of nitrogen abundances (see \ref{subsec:nitrogen}). As the goal of this project is to compare the abundances of molecules detected in the atmosphere of TOI-270~d (i.e., H$_2$O, CH$_4$ and CO$_2$) to the output of our model rather than to predict the abundances of nitrogen-bearing species, we include these models primarily as a check that the nitrogen depletion does not alter the compositions of the detected species nor introduce any additional detectable species. We discuss possible implications of the non-detections of nitrogen-bearing species in the atmosphere of TOI-270~d in further detail in Section \ref{subsec:nitrogen}.

\begin{table}
    \centering
    \setlength{\arrayrulewidth}{1.3pt}
    \begin{tabular}{cc}
    	\hline
		Parameter (unit) & Range  \\
		\hline
        $T_{\rm m-a}$ (K) & 2000--3000 \\
        $T_{\rm c-m}$ (K) & 3000--4000 \\
        $x_{\rm Fe}$ (\%) & 1--50 \\
        $K_{zz}$ (cm$^2\,$s$^{-1}$) & 10$^3$--10$^9$ \\
        \hline
    \end{tabular}
    \caption{Ranges of values considered for parameters which are varied in this study. $T_{\rm m-a}$, $T_{\rm c-m}$ and $x_{\rm Fe}$ are inputs to the core-mantle-atmosphere equilibrium chemistry model \citep{Schlichting2022}, and $K_{zz}$ is an input to the photochemical model \citep{photochem}.}
    \label{tab:params}
\end{table}

\begin{figure*}
    \centering
    \includegraphics[width=\linewidth]{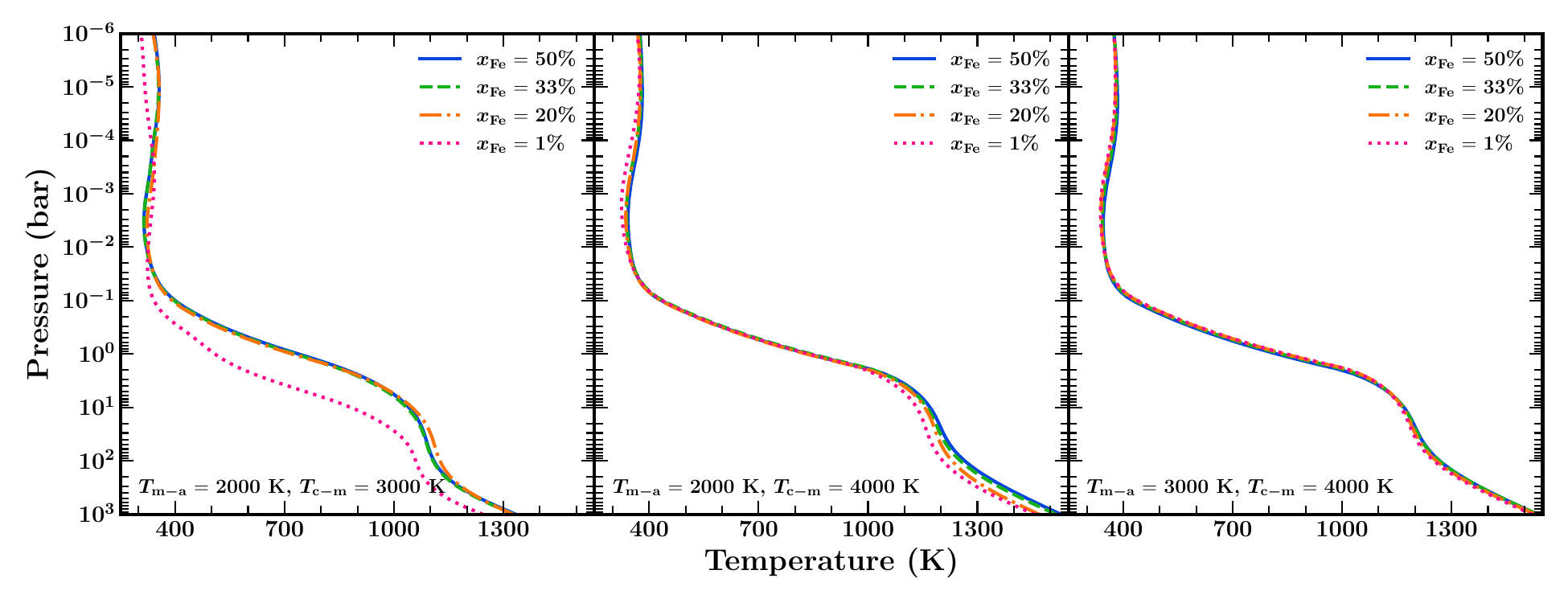}
    \caption{Temperature profiles for TOI-270~d from \texttt{HELIOS}. Each panel shows a different combination of core-mantle and mantle-atmosphere boundary temperatures. The different colors and linestyles for each profile represent different iron mass fractions, shown as a percentage of the mass of the nucleus of the planet (i.e. the iron and silicate components).}
    \label{fig:pt}
\end{figure*}

\subsection{Radiative-convective equilibrium} \label{subsec:rc}

We model atmospheric temperature profiles and emission spectra using the one-dimensional radiative transfer code \texttt{HELIOS} \citep{Malik2017,Malik2019_rocky,Malik2019}. \texttt{HELIOS} computes atmospheric temperature profiles in radiative-convective equilibrium. In this work, we assume perfect heat redistribution between the planet's two hemispheres. To achieve this, we assume a heat redistribution factor of 0.25, which dilutes the incoming radiation in order to approximate the effect of averaging over the true non-uniform irradiation pattern \citep{Hansen2008}. We include convective adjustment, with the adiabatic coefficient $\kappa$ set to 2/7, where  $\kappa = (d \ln T/d \ln P)_S$. This is appropriate for a H$_2$-rich atmosphere in the absence of significant condensation.

Our model atmosphere extends from 10$^{-6} - 10^3$~bar. We note that this model is representative of the upper atmosphere alone, with the pressure at the base of the atmosphere being at much higher pressures than 10$^3$~bar. Indeed, the pressure at the base of the atmosphere depends on the redox state of the planet and therefore on the magma-atmosphere interaction itself \citep[][]{Nicholls2024}. We adopt planetary parameters similar to those of TOI-270~d \citep{VanEylen2021,MikalEvans2023}: $\log_{10} g_p$ (cgs) $= 2.96$, $a=0.0733$~AU, $R_p = 2.19\,R_{\oplus}$. We simulate a stellar spectrum similar to TOI-270 by interpolating from a PHOENIX model grid \citep{Husser2013}, with $\log_{10} g_*$ (cgs) $= 4.872$, $T_{\rm eff} = 3506$~K, and [Fe/H] $ = -0.2$~dex.

For a given composition, we generate opacity tables using the $k$-distribution method at a resolution $R=300$ between 0.24 and 500~$\mu$m. We include opacity from the following species: C$_2$H$_2$ \citep{Chubb2020}, CH$_4$ \citep{Hargreaves2020}, CO, CO$_2$ \citep{Rothman2010}, H$_2$O \citep{Polyansky2018}, H$_2$S \citep{Azzam2016,Chubb2018}, HCN \citep{Barber2014}, NH$_3$ \citep{Coles2019}, O$_2$ \citep{HITRAN2020}, PH$_3$ \citep{SousaSilva2015}, SiH$_4$ \citep{Owens2017}, SiO \citep{Barton2013}, Fe, K, Mg and Na \citep{Kurucz1995}, as well as H$_2$-H$_2$ and H$_2$-He collision-induced absorption \citep{Richard2012} and Rayleigh scattering. Each of these opacity sources covers a wide temperature range, meaning no extrapolation is required to apply them in our model. Pressure broadening is applied following \citep{Grimm2021} to cover the full pressure range of the models.

The output of \textsc{fastchem cond} is required to produce opacity tables which are input to \texttt{HELIOS}. However, it is not possible to account for rainout condensation without an initial \texttt{HELIOS} model to provide a pressure-temperature profile. We therefore initially assume equilibrium chemistry with no condensation across a wide range of temperatures (100-6000~K) which we input to \texttt{HELIOS} to produce an initial guess for the pressure-temperature profile. We then use this pressure-temperature profile as an input for \textsc{fastchem cond} with rainout condensation included. This produces new chemical abundance profiles, which can then be returned to \texttt{HELIOS}. We iterate between the two models until the pressure-temperature profile converges. Our convergence criterion is that the difference between subsequent temperature profiles does not exceed 20~K in any layer of the atmosphere. Our initial \texttt{HELIOS} model without condensation uses a pre-mixed opacity table. Subsequent iterations in which condensation is accounted for use on-the-fly opacity mixing using the random overlap resort-rebin method \citep{Amundsen2017}.

\begin{figure*}
    \centering
    \includegraphics[width=\linewidth]{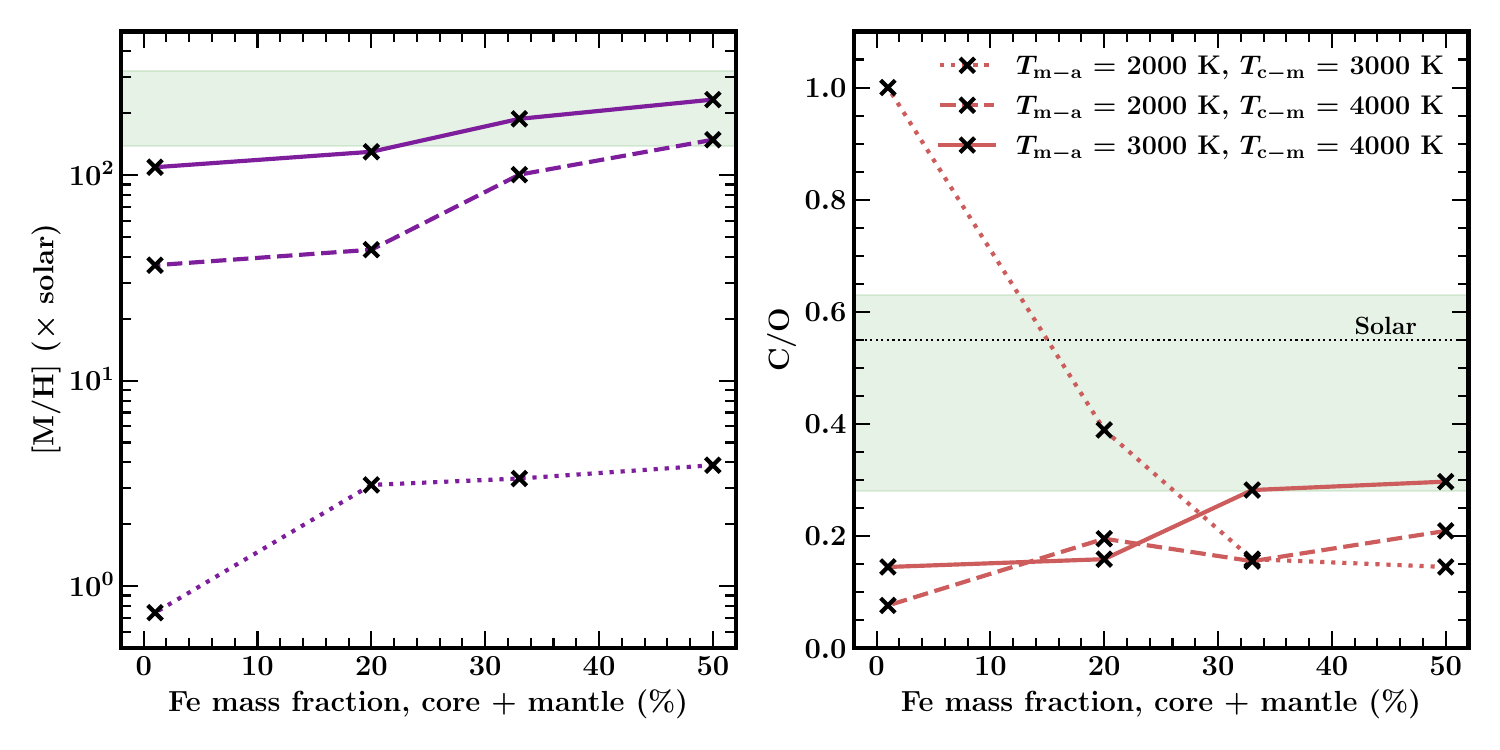}
    \caption{Metallicity ([M/H], left) and C/O ratio (right) of the upper atmosphere of TOI-270~d ($P=1\,$mbar) for different magma ocean model scenarios. [M/H] and C/O are plotted as a function of the iron mass fraction, shown as a percentage of the total mass of the nucleus (core + mantle). The metallicity [M/H] is calculated as [(C+O)/H] relative to solar abundances \citep{Asplund2009}, since carbon and oxygen-bearing species are the main drivers of the metallicity constraints presented in \citet{Benneke2024}. Different line styles represent different combinations of mantle-atmosphere and core-mantle boundary temperatures ($T_{\rm m-a}$ and $T_{\rm c-m}$ respectively). The green shaded regions show the constraints on [M/H] and C/O reported by \citet{Benneke2024}, inferred from the JWST transmission spectrum of the planet. The black dotted line on the right-hand panel shows the solar C/O value from \citet{Asplund2009}.}
    \label{fig:met_c_to_o}
\end{figure*}

\subsection{Photochemistry}

Photochemical processes and vertical mixing also play a role in altering the chemical composition of the upper atmosphere. We account for these effects using the \textit{Photochem} package \citep{Wogan2023,photochem}. \textit{Photochem} is a chemical disequilibrium kinetics code which models the effects of thermochemistry, photolysis and mixing, with elements inherited from \citet{Zahnle2016}. We use the temperature profile and chemical abundances computed in previous steps as inputs to the photochemical model. \textit{Photochem} performs calculations on an altitude grid that is initially derived initially from these inputs. Because disequilibrium chemistry and mixing introduce deviations in mean molecular weight at each altitude grid point, \textit{Photochem} continually alters and re-grids the altitude-pressure profile such that the temperature profile is conserved \citep{Mukherjee2025}. We adopt the same planetary and stellar parameters as the previous modeling steps, apart from the stellar spectrum (see below). There is no strong evidence for hazes in the atmosphere of TOI-270~d, so we do not include aerosol particles in these models. We vary the strength of vertical mixing via the eddy diffusion coefficient $K_{zz}$, ranging from 10$^3$--10$^9\,$cm$^2\,$s$^{-1}$ in steps of 1~dex. This range is chosen to follow previous studies of sub-Neptune photochemistry \cite[e.g.,][]{Tsai2021} while also allowing for lower values which have been suggested for solar system giants \citep{Moses2005}.

It is important to use an appropriate stellar spectrum to ensure accurate photochemical modeling. Since a stellar spectrum of TOI-270 of sufficiently high quality is not available, we use the spectrum of GJ~832 as a proxy. GJ~832 has the same stellar type \citep[M3V,][]{Pineda2021} as TOI-270 \citep{MikalEvans2023} and comparable effective temperature and surface gravity \citep[$T_{\rm eff}=3539^{+79}_{-74}\,$K, $\log g \, (\rm{cgs}) = 4.792^{+0.036}_{-0.039}$ for GJ~832, $T_{\rm eff}=3506\pm 70\,$K, $\log g = 4.872 \pm 0.026$ for TOI-270,][]{Pineda2021,VanEylen2021}. The spectrum is acquired from the MUSCLES survey \citep{France2016,Youngblood2016,Lloyd2016}.

\subsection{Forward model spectra}

Finally, we generate transmission spectra from the outputs of the previous modeling steps using \textsc{Aura-3D} \citep{Nixon2022}. \textsc{Aura-3D} includes a radiative transfer scheme that allows for the generation of spectra with thermal and chemical profiles that vary with height, longitude and latitude in the atmosphere. It is built on the \textsc{Aura} family of retrieval codes \citep{Pinhas2018,Nixon2020}, with the version used in this project also incorporating developments presented in \citet{Welbanks2021} and \citet{Nixon2024_unc}. Although \textsc{Aura-3D} can generate models with three-dimensional geometry, temperature and abundance profiles vary with height only in this work, since the outputs of \texttt{HELIOS} and \textit{Photochem} only depend on height (i.e. pressure).

\textsc{Aura-3D} is used to compute the transmission spectrum for a transiting planet, accounting for opacity contributions from chemical species, collision-induced absorption (CIA), and clouds or hazes. We use the same line lists as described in Section \ref{subsec:rc}, and use opacity sampling at $R=60\,000$ to generate forward models. Volume mixing ratios and pressure-temperature profiles are taken directly from the output of \textit{Photochem} and \texttt{HELIOS} respectively. We note that some gases (e.g., N$_2$) included in the \textit{Photochem} output are not spectrally active, so their opacities are not included in the forward model. However, their volume mixing ratios are still accounted for in order to calculate the atmospheric mean molecular weight, which plays a role in determining the amplitude of atmospheric features in the transmission spectrum. In keeping with other components of our modeling framework, we adopt planetary parameters of TOI-270~d and stellar parameters of TOI-270 from \citet{VanEylen2021} and \citet{MikalEvans2023}.

\begin{figure*}
    \centering
    \includegraphics[width=\linewidth]{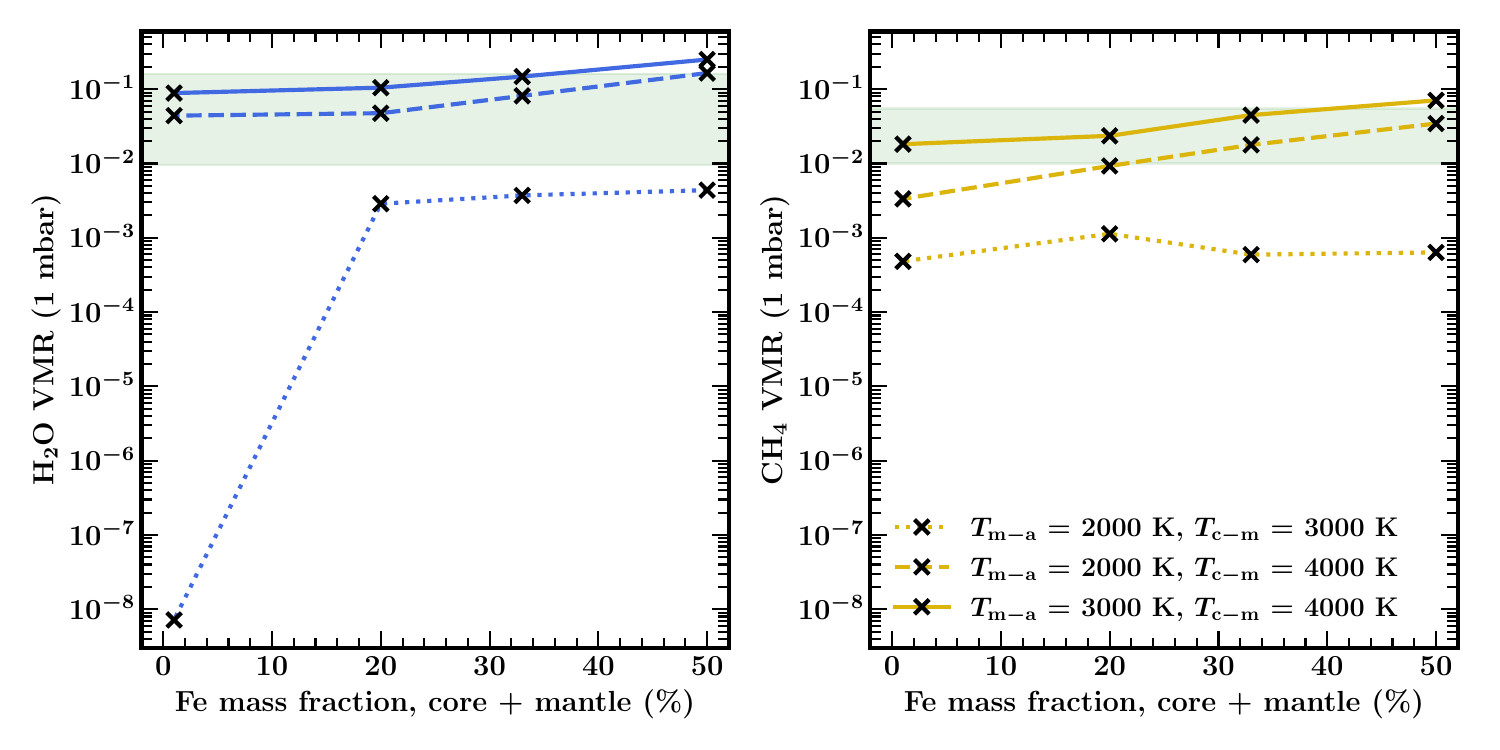}
    \caption{Volume mixing ratio of H$_2$O (left) and CH$_4$ (right) in the upper atmosphere of TOI-270~d ($P=1\,$mbar) for different magma ocean model scenarios. We show models with $K_{zz}=10^7$cm$^2$s$^{-1}$, although the difference in H$_2$O and CH$_4$ volume mixing ratio is negligible for different values of $K_{zz}$. Volume mixing ratios are plotted as a function of the iron mass fraction, shown as a percentage of the total mass of the nucleus (core + mantle). Different line styles represent different combinations of mantle-atmosphere and core-mantle boundary temperatures ($T_{\rm m-a}$ and $T_{\rm c-m}$ respectively). The green shaded regions show the constraints on the volume mixing ratios of H$_2$O and CH$_4$ reported by \citet{Benneke2024}, inferred from the JWST transmission spectrum of the planet.}
    \label{fig:h2o_ch4}
\end{figure*}

\section{Results}\label{sec:results}

In this section we describe the vertical atmospheric temperature profiles and compositions that are produced by our modeling framework, and compare our results to the retrieved atmospheric composition of TOI-270~d found by \citet{Benneke2024}.

\subsection{Temperature profiles}

Figure \ref{fig:pt} shows converged atmospheric temperature profiles from \texttt{HELIOS} for the range of model parameters considered. All temperature profiles converged within 2 iterations between \texttt{HELIOS} and \textsc{fastchem cond}, although there were some more significant deviations ($\Delta T>150$~K) between the initial temperature profile without condensation and the first temperature profile calculated once condensation had been accounted for. Overall, there is not significant variation in the converged temperature profiles for different scenarios. We find that higher core-mantle and mantle-atmosphere boundary temperatures lead to hotter temperatures in the deep atmosphere. Models with higher iron mass fractions also have hotter temperature profiles, although the effect becomes less pronounced for models with hotter boundary temperatures. The difference is most notable at pressures higher than 1~bar. This is likely a result of models with hotter boundary temperatures having higher atmospheric metallicity, including higher H$_2$O and CH$_4$ abundances (see Figure \ref{fig:h2o_ch4}). Higher H$_2$O and CH$_4$ abundances cause the atmosphere to be optically thick at lower pressures, which results in higher temperatures in the deep atmosphere due to a stronger greenhouse effect \citep[e.g.,][]{Kempton2023_ww}. We note that the mantle-atmosphere boundary temperature is not equivalent to the temperature at 10$^3$ bar---the mantle-atmosphere boundary is found at higher pressures than those modeled by \texttt{HELIOS}.

\subsection{Atmospheric metallicity and C/O}

We begin our exploration of the resulting atmospheric chemistry by investigating the metallicity and C/O of the upper atmosphere ($P=1\,$mbar), as depicted in Figure \ref{fig:met_c_to_o}. Since the metallicity constraint for TOI-270~d is derived from the volume mixing ratios of carbon- and oxygen-bearing species, we use [(C+O)/H] as a proxy for metallicity. For most of the models considered in this work, magma ocean interactions lead to an atmospheric metallicity that is enhanced relative to solar, and a C/O that is depleted relative to solar. This qualitatively agrees with previous work exploring magma-atmosphere interactions \citep[e.g.,][]{Seo2024,Werlen2025}. The extent of these changes to metallicity and C/O depend on the thermodynamic conditions at the atmosphere-mantle and mantle-core boundaries, as well as the iron mass fraction in the nucleus of the planet. The atmospheric metallicity increases as the boundary temperatures increase, and also increases as the iron mass fraction increases. One of the most important reaction pathways leading to the formation of water requires participation of elements found in the iron-rich core \citep{Kite2020,Schlichting2022}, which explains why metallicity decreases when the core mass fraction decreases.

A notable exception to the overall finding that metallicity is enhanced and C/O is depleted is the case where $T_{\rm m-a}=2000\,$K, $T_{\rm c-m}=3000\,$K and $x_{\rm Fe}=1\%$. For these parameters, the atmospheric metallicity becomes slightly subsolar, and the C/O is enhanced to $\sim$1. These effects are explained in part due to the low iron mass fraction, as described above, and also as a result of the cooler temperature profile for this model, which leads to H$_2$O condensation (see Section \ref{subsec:abundances}). A low atmospheric metallicity could therefore be the result of an interior with very little chemically-reactive iron. However, it is unclear how such an iron-poor planet would form \citep{Rogers2015}. Additionally, flotation or suspension of metal droplets in the convective magma ocean have been suggested to increase the mass of chemically-reactive iron in the magma oceans of sub-Neptunes \citep{Lichtenberg2021,Young2024}. We further note that in this case, the assumption that the adiabatic index $\kappa=2/7$ may no longer be appropriate, due to the role of condensation. We reserve further examination of the impact of H$_2$O condensation on these models for a future study, since there is no evidence for H$_2$O depletion in the upper atmosphere of TOI-270~d.

The green shaded regions in Figure \ref{fig:met_c_to_o} show the retrieved metallicity and C/O of the atmosphere of TOI-270~d from \citet{Benneke2024}. The retrieved atmospheric metallicity from that work is 225$^{+98}_{-86}\times$ solar, and the retrieved C/O is 0.47$^{+0.16}_{-0.19}$. We find that two of our models are consistent with both values to within 1$\sigma$. This is achieved when $T_{\rm m-a}=3000\,$K, $T_{\rm c-m}=4000\,$K and $x_{\rm Fe}=33\%$ or $50\%$. This suggests that magma-atmosphere interactions may be able to explain the observed atmosphere of the planet. Our next step is to move beyond atmospheric metallicity and C/O, and compare the abundances of specific chemical species that were detected in the planet's atmosphere \citep{Benneke2024} to the outputs of our model.

\subsection{Molecular volume mixing ratios} \label{subsec:abundances}

The resulting volume mixing ratios of H$_2$O and CH$_4$ at 1~mbar are shown in Figure \ref{fig:h2o_ch4}. Similarly to the atmospheric metallicity, we find that H$_2$O and CH$_4$ are enhanced relative to solar composition for both scenarios in which $T_{\rm c-m}=4000\,$K \citep[where solar $\log_{10}X_{\rm{H_2O}}=-3.0,\,\log_{10}X_{\rm{CH_4}}=-3.3$, e.g.][]{Moses2013b}. This enhancement increases further as $T_{\rm m-a}$ rises. When $T_{\rm c-m}=3000\,$K, the H$_2$O volume mixing ratio is only slightly to increased relative solar composition, while the CH$_4$ volume mixing ratio remains approximately solar. As seen in Figure \ref{fig:met_c_to_o}, the model with the lowest iron mass fraction is an outlier, with the H$_2$O volume mixing ratio depleted due to condensation.

We find that numerous combinations of model parameters can explain the retrieved H$_2$O and CH$_4$ volume mixing ratios for TOI-270~d. \cite{Benneke2024} report $\log_{10}X_{\rm{H_2O}}=-1.10^{+0.31}_{-0.92},\,\log_{10}X_{\rm{CH_4}}=-1.64^{+0.38}_{-0.36}$. Seven of the twelve models shown in Figure \ref{fig:h2o_ch4} yield H$_2$O volume mixing ratios that are consistent with the retrieved H$_2$O, and five yield CH$_4$ volume mixing ratios that are consistent with the retrieved CH$_4$. This indicates that a magma-atmosphere interaction model can readily reproduce these volume mixing ratios.

\begin{figure}
    \centering
    \includegraphics[width=\linewidth]{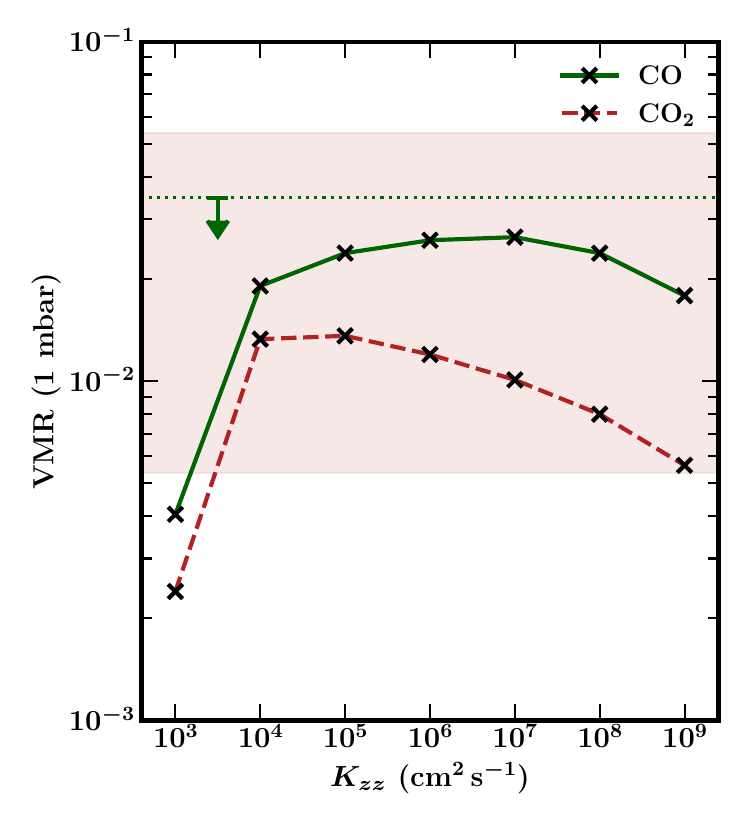}
    \caption{Volume mixing ratios of CO (green, solid) and CO$_2$ (red, dashed) in the upper atmosphere of TOI-270~d ($P=1\,$mbar) as a function of $K_{zz}$. $T_{\rm{m-a}}=3000\,$K, $T_{\rm{c-m}}=4000\,$K and $x_{\rm Fe}=33\%$ for all models shown here. The red shaded region shows the constraints on the volume mixing ratio of CO$_2$ reported by \citet{Benneke2024}, and the green dotted line shows the reported 2$\sigma$ upper limit for CO.}
    \label{fig:co_co2}
\end{figure}

We find that the volume mixing ratios of H$_2$O and CH$_4$ are largely unaffected by vertical mixing and photochemistry. For this reason, Figure \ref{fig:h2o_ch4} only shows results for $K_{zz}=10^7$cm$^2$s$^{-1}$, with other values yielding essentially the same values. The same cannot be said for the volume mixing ratios of CO and CO$_2$. Without considering vertical mixing, the volume mixing ratios of CO and CO$_2$ sharply decrease at $P\lesssim1\,$bar, which is inconsistent with the detection of CO$_2$ reported by \citep{Benneke2024}. However, vertical mixing serves to enhance the upper atmospheric concentrations of these molecules considerably, a phenomenon which has been reported in models of other planets \citep[e.g., WASP-107~b,][]{Welbanks2024}.

Figure \ref{fig:co_co2} compares the volume mixing ratios of CO and CO$_2$ at 1~mbar for models with $T_{\rm{m-a}}=3000\,$K, $T_{\rm{c-m}}=4000\,$K and $x_{\rm Fe}=33\%$ across the range of values of $K_{zz}$ considered in our photochemical model. We choose to highlight this set of parameters since they yield values of [M/H], C/O, $X_{\rm H_2O}$ and $X_{\rm CH_4}$ that are consistent with atmospheric observations of TOI-270~d. The CO and CO$_2$ concentrations above the quench pressure are homogenized to their values at the quench pressure by vertical mixing. As $K_{zz}$ increases, the quench pressure moves deeper in the atmosphere, and the volume mixing ratios change accordingly. We find that for $K_{zz}\geq$10$^{4}$, the volume mixing ratio of CO$_2$ is consistent with the retrieved $\log_{10}X_{\rm{CO_2}}=-1.67^{+0.40}_{-0.60}$ from \citet{Benneke2024}. We find that the volume mixing ratio of CO is also enhanced, but remains below the 2$\sigma$ upper limit reported in that work ($\log_{10}X_{\rm{CO}}<-1.46$). Such an increase of CO over CO$_2$ is consistent with \citet{Shorttle2024} and \citet{Nicholls2024}, who identified increasing atmospheric CO for intermediate magma ocean redox states as an important tracer of interaction with the interior.

The bottom panel of Figure \ref{fig:abundances} shows the volume mixing ratios of species at the atmosphere-mantle boundary, i.e. the output of the \citet{Schlichting2022} model. We note that several species appear at the atmosphere-mantle boundary which are not present in the upper atmosphere, including the silicon-bearing species SiO and SiH$_4$. These species have previously been highlighted as possible signatures of atmosphere-interior interactions \citep[e.g.,][]{Misener2023}. However, we do not expect these species to be present in the upper atmosphere of TOI-270~d due to condensation. We confirm this by comparing our final model to our initial model that does not account for condensation, and find that in the case without condensation, SiO and SiH$_4$ are present in appreciable quantities, similar to their volume mixing ratios at the atmosphere-mantle boundary. This suggests that for hotter planets, SiO and SiH$_4$ could indeed be present and used to infer atmosphere-interior interactions, although a detailed investigation is beyond the scope of this work.

In order to test whether the uncertainty in the total iron and silicate budget of our planet impacts our results, we include two additional model cases with the total iron and silicate mass set to 3.8$\,M_{\oplus}$ and 4.8$\,M_{\oplus}$, spanning the reported $1\sigma$ range from \citet{Benneke2024}. For these models we fix $T_{\rm{m-a}}=3000\,$K, $T_{\rm{c-m}}=4000\,$K, $x_{\rm Fe}=33\%$ and $K_{zz}=10^7$cm$^2$s$^{-1}$. We find that the upper atmospheric concentrations of H$_2$O, CH$_4$, CO and CO$_2$ deviate slightly from their values in the 4.3$\,M_{\oplus}$, but by less than 0.04~dex in all cases, much smaller than the reported uncertainty on measurements from JWST. We therefore conclude that uncertainty in the iron and silicate mass does not alter our overall findings.

\subsection{Comparison with JWST observations}

We find that a subset of our models are able to match the retrieved atmospheric abundances for TOI-270~d reported in \citet{Benneke2024}. Figure \ref{fig:abundances} shows the volume mixing ratios of key chemical species as a function of pressure in the case where $T_{\rm{m-a}}=3000\,$K, $T_{\rm{c-m}}=4000\,$K, $x_{\rm Fe}=33\%$ and $K_{zz}=10^7$cm$^2$s$^{-1}$. We find that all of the major carbon- and oxygen-bearing species are consistent within the 1$\sigma$ retrieved abundances for species which were detected in the atmosphere (i.e., H$_2$O, CH$_4$ and CO$_2$), or below the reported upper limit in the case that the species was not detected (i.e., CO).

We further demonstrate the capability of our model to explain JWST observations of TOI-270~d by generating a forward model transmission spectrum for direct comparison to the data. We use the temperature profile and chemical abundances from our model with $T_{\rm m-a}=3000\,$K, $T_{\rm c-m}=4000\,$K, $x_{\rm Fe}=33\%$ and $K_{zz}=10^7$cm$^2$s$^{-1}$. Additionally, we varied the nitrogen abundance to account for its possible depletion (see Section \ref{subsec:nitrogen}). Two more parameters are required to produce a forward model spectrum: a reference pressure $P_{\rm ref}$ and cloud deck pressure $P_{\rm cloud}$. $P_{\rm ref}$ is the pressure assigned to the white-light radius of the planet and has the effect of changing the transit depth by a constant value across all wavelengths, whereas $P_{\rm cloud}$ reduces the size of atmospheric spectral features by setting the optical depth to $\infty$ for $P>P_{\rm cloud}$. \citet{Benneke2024} reported a lower limit on $P_{\rm cloud}$ of 10$^{-2.99}$ bar, and did not report constraints on $P_{\rm ref}$. We therefore generated a grid of models with $P_{\rm cloud}>10^{-2.99}$ bar and $P_{\rm ref}$ within the full range of pressures from the chemistry models, with a step size of 0.5 dex. This allowed us to find a best-fit spectrum given our chosen chemistry and temperature profiles by selecting the model that resulted in the lowest $\chi^2/n_{\rm data}$. When fitting the NIRISS/SOSS and NIRSpec/G395H data, we assumed an offset of 12 ppm between the spectra of each instrument, according to the offset reported by \citet{Benneke2024}.

Our resulting spectra are shown in Figure \ref{fig:spectrum}. Our optimal parameter values were $P_{\rm ref}=1$~bar, $P_{\rm cloud}=10^{-2.5}$~bar. This parameter combination yielded $\chi^2/n_{\rm data}=1.18$ for the model with nitrogen depleted to 10$^{-2}\times\,$solar. For this dataset, we can approximate the standard deviation of $\chi^2/n_{\nu}$ to be 0.13, following \citet{Andrae2010}. This indicates that models with $0.74 \leq \chi^2/n_{\nu} \leq 1.26$ are consistent with the data at the 2$\sigma$ level. The value of $\chi^2/n_{\nu}$ is likely somewhat higher than our calculated $\chi^2/n_{\rm data}=1.18$, though it is difficult to estimate the number of degrees of freedom, $n_{\nu}$, for a non-linear model. Nevertheless, these calculations suggest that the model agrees with the data at the $<$2$\sigma$ level. We deem this to be a good level of agreement, particularly given that this is a self-consistent forward model rather than a retrieval result from a full exploration of parameter space. We note that spectra with 1$\times\,$solar and 10$^{-4}\times\,$solar nitrogen abundances are almost indistinguishable from the 10$^{-2}\times\,$solar case. We discuss the scenario with enhanced nitrogen relative to solar values in Section \ref{subsec:nitrogen}.

\citet{Holmberg2024} also presented an analysis of the HST/WFC3 + NIRSpec/G395H transmission spectrum of TOI-270~d. That study reported a range of abundance constraints depending on factors such as the treatment of limb darkening. A number of our models with lower iron mass fractions and/or $T_{\rm c-m}=3000\,$K come close to matching their reported abundances. For example, the ``one offset + constant limb darkening'' analysis presented by \citet{Holmberg2024} reports the following abundances: $\log_{10}X_{\rm{H_2O}}=-1.04^{+0.24}_{-0.45}$, $\log_{10}X_{\rm{CH_4}}=-2.97^{+0.30}_{-0.39}$, $\log_{10}X_{\rm{CO_2}}=-3.95^{+0.72}_{-0.90}$ and $\log_{10}X_{\rm{CO}}<-3.37$ (95\% upper limit). By comparison, our model where $T_{\rm{m-a}}=2000\,$K, $T_{\rm{c-m}}=4000\,$K, $x_{\rm Fe}=1\%$ and $K_{zz}=10^6\,$cm$^2\,$s$^{-1}$ has $\log_{10}X_{\rm{H_2O}} = -1.36$, $\log_{10}X_{\rm{CH_4}}=-2.48$, $\log_{10}X_{\rm{CO_2}}=-4.73$ and $\log_{10}X_{\rm{CO}}=-3.87$. Each of these volume mixing ratios agree with the measured values, with the exception of CH$_4$, which we find to have a slightly higher concentration than the 1$\sigma$ range reported by \citet{Holmberg2024}. We discuss this finding in more detail in Section \ref{subsec:required}.


\subsection{Impact of nitrogen depletion} \label{subsec:nitrogen}
\begin{figure}
    \centering
    \includegraphics[width=\linewidth]{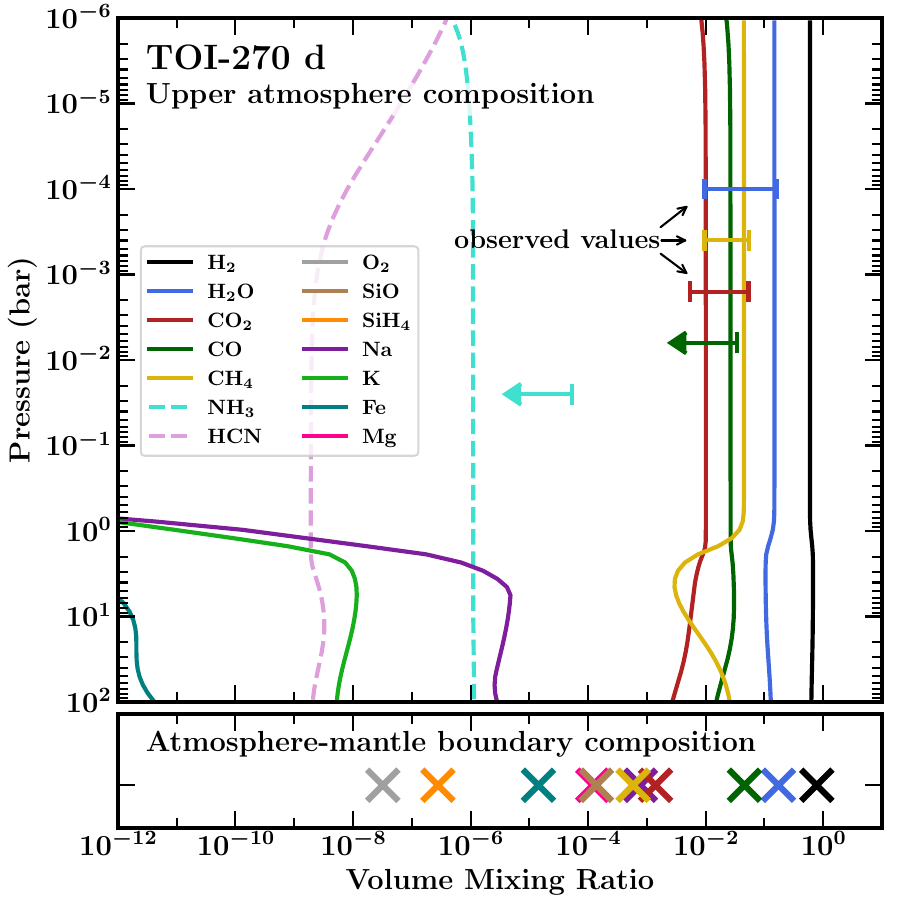}
    \caption{Volume mixing ratio profiles for chemical species in the atmosphere of TOI-270~d, based on interactions between a magma ocean and a gaseous atmosphere with an initial solar composition. Model parameters are $T_{\rm m-a}=3000\,$K, $T_{\rm c-m}=4000\,$K, $x_{\rm Fe}=33\%$, $K_{zz}=10^7\,$cm$^2$s$^{-1}$. The bottom panel shows volume mixing ratios of species at the mantle-atmosphere interface ($P \gg 100\,$bar). Error bars and arrows show measured volume mixing ratios or upper limits for H$_2$O, CH$_4$ and CO$_2$, CO and NH$_3$ in the upper atmosphere, derived from JWST observations of TOI-270~d \citep{Benneke2024}. Dashed lines show nitrogen-bearing chemical species not included in the magma ocean model, but included in the upper atmosphere model. In the case shown here, atmospheric nitrogen is depleted to 10$^{-2}\times\,$solar.}
    \label{fig:abundances}
\end{figure}

Since the model used to calculate chemical equilibrium abundances between the core, mantle and atmosphere does not include nitrogen- or sulfur-bearing species, our work primarily focuses on determining the abundances of carbon- and oxygen-bearing species that were detected in the atmosphere of TOI-270~d. However, \citet{Shorttle2024} demonstrated that magma-atmosphere interactions could lead to depleted atmospheric nitrogen, impacting the abundances of species such as NH$_3$. Although we leave the full incorporation of nitrogen chemistry into our model for a further study, we conduct a sensitivity test to determine whether the abundances of other atmospheric species are impacted by varying the atmospheric nitrogen abundances. We also explore the extent to which the upper limit for NH$_3$ reported by \citet{Benneke2024} can inform us as to whether nitrogen is depleted in the atmosphere of TOI-270~d.

\begin{figure*}
    \centering
    \includegraphics[width=\textwidth]{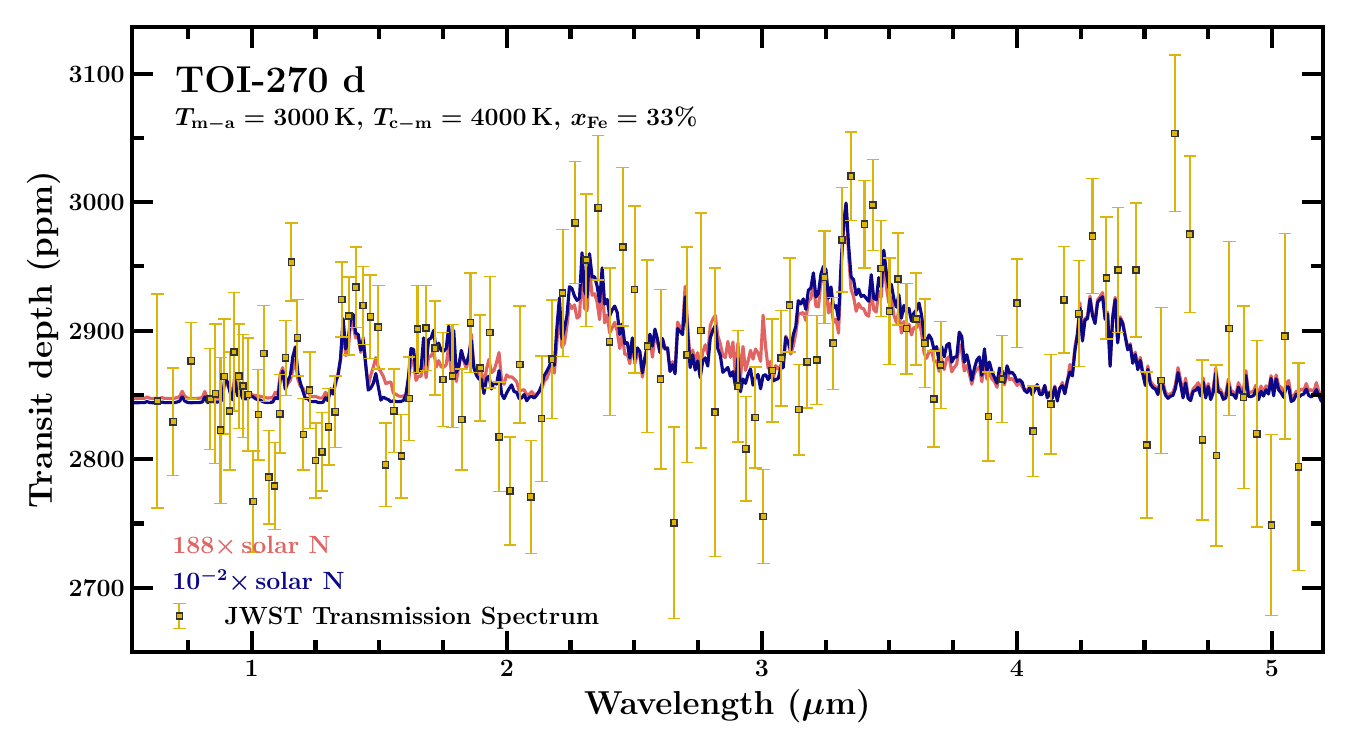}
    \caption{Transmission spectrum of TOI-270~d. Gold squares with error bars show the observed JWST spectrum from \citet{Benneke2024}. The pink and blue lines show forward modeled spectra assuming the upper atmospheric composition from our coupled magma-atmosphere model. Model parameters are $T_{\rm m-a}=3000\,$K, $T_{\rm c-m}=4000\,$K, $x_{\rm Fe}=33\%$, $K_{zz}=10^7\,$cm$^2$s$^{-1}$. The pink line shows a case where nitrogen is enhanced in line with the carbon and oxygen enhancement found by the magma-atmosphere model. The blue line shows a case where nitrogen is depleted to 10$^{-2}\times\,$solar abundance, in line with expectations of nitrogen depletion from magma-atmosphere interactions \citep{Shorttle2024}. For the model with enhanced nitrogen, $\chi^2/n_{\rm data}=1.31$, and for the model with depleted nitrogen, $\chi^2/n_{\rm data}=1.18$.}
    \label{fig:spectrum}
\end{figure*}

For our sensitivity test, we focus on models where $T_{\rm{m-a}}=3000\,$K, $T_{\rm{c-m}}=4000\,$K and $x_{\rm Fe}=33\%$. These parameters are chosen since they provide the best overall explanation for the observed atmosphere of TOI-270~d. For the photochemistry and vertical mixing, we include the full range of $K_{zz}$ values included in our previous model grid. We wish to determine whether varying the nitrogen abundance leads to changes in the volume mixing ratios of the three chemical species that were detected in the atmosphere of TOI-270~d: H$_2$O, CH$_4$ and CO$_2$. Additionally, we compare the CO volume mixing ratio to its measured upper limit. We consider four different values for the nitrogen abundance by volume: 188$\times$ solar, 1$\times$ solar, $10^{-2}\times$ solar and $10^{-4}\times$ solar. Our maximum value is chosen to match the atmospheric metallicity [(C+O)/H] of the magma-atmosphere model. We include a 1$\times$ solar case to represent a scenario in which nitrogen is unaffected by magma-atmosphere interactions. The remaining two cases reflect scenarios in which nitrogen is depleted from the atmosphere, with values spanning the range presented by \citet{Shorttle2024}.

For the 1$\times$ solar, $10^{-2}\times$ solar and $10^{-4}\times$ solar cases, we find that changes in the abundances of H$_2$O, CH$_4$, CO and CO$_2$ are negligible. Across all chemical species and $K_{zz}$ values, the maximum fractional change in abundance at 1~mbar is $\Delta X_i/X_i = 6.4\times10^{-4}$. We therefore conclude that our results are not sensitive to possible depletion of nitrogen in the atmosphere. However, we do find that the abundances of these species change when we allow for nitrogen to be enhanced to the metallicity indicated by the C and O abundances. In this case, we see an increase in the H$_2$O volume mixing ratio and a decrease in CH$_4$, CO and CO$_2$. For example, when $K_{zz} = 10^7$~cm$^2\,$s$^{-1}$, the volume mixing ratios of  CH$_4$, CO and CO$_2$ decrease by a factor of 2.35, 1.73 and 1.36 respectively in the enhanced nitrogen model. For these species, the depleted values are still consistent with the abundance estimates from \citet{Benneke2024}. However, the H$_2$O volume mixing ratio increases from 14.8\% in the solar nitrogen case to 18.1\% in the enhanced nitrogen case, making it slightly higher than the 1$\sigma$ upper limit of 16.2\% reported by \citet{Benneke2024}. This suggests that an atmosphere uniformly enhanced in metals relative to hydrogen may be unlikely for the target, potentially lending further evidence that its atmosphere has been affected by  magma ocean interactions, although we note that the change in goodness-of-fit to observations between these models is quite small (see Figure \ref{fig:spectrum} and Section \ref{subsec:nh3}).

\begin{figure}[t]
    \centering
    \includegraphics[width=\linewidth]{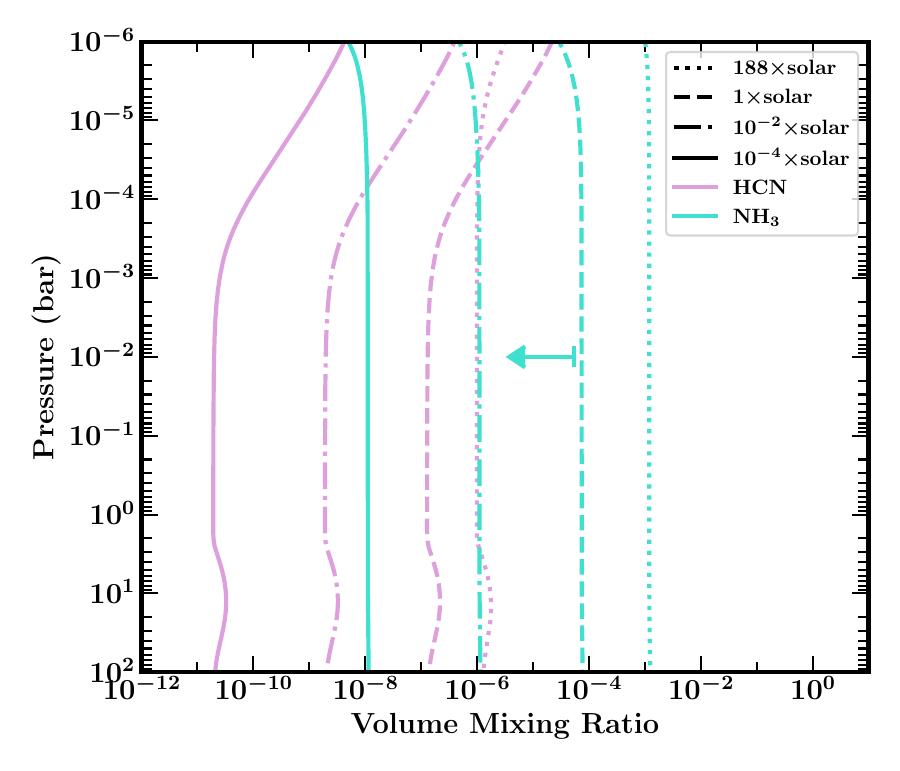}
    \caption{Volume mixing ratio profiles for NH$_3$ and HCN for a range of initial atmospheric nitrogen abundances. Model parameters are $T_{\rm m-a}=3000\,$K, $T_{\rm c-m}=4000\,$K, $x_{\rm Fe}=33\%$, $K_{zz}=10^7\,$cm$^2$s$^{-1}$. The 188$\times$ solar case represents nitrogen enhancement in line with the derived metallicity from the resulting carbon and oxygen abundances from the magma-atmosphere model. The arrow represents the 2$\sigma$ upper limit for the NH$_3$ volume mixing ratio according to \citet{Benneke2024}.}
    \label{fig:nh3_hcn}
\end{figure}

We also explored how varying the total nitrogen budget for the atmosphere impacts the presence of key nitrogen-bearing species in the atmosphere, namely NH$_3$ and HCN. The volume mixing ratios of these two species for different levels of nitrogen enhancement are shown in Figure \ref{fig:nh3_hcn}. We focus on models where $K_{zz} = 10^7$~cm$^2\,$s$^{-1}$, as the volume mixing ratios of other species are consistent with observations in this case (see Figure \ref{fig:abundances}). This figure demonstrates how the NH$_3$ abundance decreases as the total amount of nitrogen in the atmosphere decreases. We find that models assuming $10^{-2}\times$ solar and $10^{-4}\times$ solar nitrogen abundances are consistent with the upper limits for NH$_3$ presented by \citet{Benneke2024}, whereas the 188$\times$ solar and 1$\times$ solar models exceed this upper limit, albeit only slightly for the 1$\times$ solar case. This implies that nitrogen is likely to be depleted in this atmosphere, which is qualitatively consistent with magma ocean interactions according to \citet{Shorttle2024}. However, we note that detecting and constraining the abundance of NH$_3$ is challenging for this target (see Section \ref{subsec:nh3} for further detail).

\section{Discussion and Conclusions}\label{sec:discussion}

\subsection{Does TOI-270~d require a magma ocean to explain its observed atmosphere?} \label{subsec:required}

Our findings suggest that the observed chemical abundances in the atmosphere of TOI-270~d can be readily explained as the outcome of interactions between an initially H$_2$-dominated atmosphere and an underlying magma ocean. This suggests that the planet need not have formed beyond the snowline, where accretion of icy material is more prevalent, to explain its high atmospheric metallicity and water content. However, while magma-atmosphere interactions may be \textit{sufficient} to explain the observed atmosphere, it is not yet possible to claim that they are \textit{necessary}. Indeed, accretion of ices at formation is still a possible cause of this planet's atmospheric composition. Further work will be required to distinguish between different formation and evolutionary scenarios. In particular, assessing the outcome of interactions between a hydrogen atmosphere and a mixed rock/ice layer would be an interesting avenue for further study. However, even after accounting for different compositions of the core and mantle, evolutionary processes such as photoevaporative atmospheric escape and core-powered mass loss could further alter the atmospheric composition \citep{Schulik2023,Cherubim2025,Heng2025} over similar or longer timescales compared to magma-atmosphere interactions. In order to truly assess the prevalence of magma ocean interactions on sub-Neptunes, a more complete evolutionary model is required. However, we believe that this work, alongside similar studies \citep{Shorttle2024,Nicholls2025}, highlights that developing an understanding of how magma oceans shape sub-Neptune atmospheres will be critical as we strive to better understand the origins and nature of these objects.

In contrast to \citet{Benneke2024}, who suggest that their findings are consistent with TOI-270~d hosting a mixed supercritical atmosphere atop an iron+rock nucleus, \citet{Holmberg2024} argue that the atmosphere of TOI-270~d is consistent with the planet hosting a liquid water ocean, citing detections of CO$_2$ and CH$_4$ alongside a non-detection of NH$_3$ as evidence for this scenario. Our work demonstrates that detectable levels of CO$_2$ and CH$_4$ can be present in the upper atmosphere as a result of magma-atmosphere interactions and vertical mixing. Similar results have been obtained for the cooler sub-Neptune K2-18~b, another planet with claims of atmospheric CO$_2$ and CH$_4$ \citep{Madhusudhan2023}. Both a liquid water ocean \citep{Madhusudhan2023} and a combination of magma-atmosphere interactions and vertical mixing \citep{Shorttle2024,Wogan2024} have been suggested to explain this composition. The qualitative agreement between the compositions predicted by both scenarios suggests that additional work is required to find unique chemical signatures of both magma-atmosphere interactions and liquid water oceans.

Our model grid includes several cases which closely match the abundance constraints from \citet{Benneke2024}. While some of our models also find relatively good agreement with the abundance constraints from \citet{Holmberg2024}, those models typically have lower iron mass fractions and core-mantle interface temperatures. However, we do not believe that the \citet{Holmberg2024} results therefore indicate a true preference for a silicate-rich and/or cooler interior. In this work we did not conduct a detailed sampling of the parameter space, instead generating models with a smaller set of reasonable parameter values. It is therefore possible that some additional unexplored combination of parameters may explain the \citet{Holmberg2024} results without a low $x_{\rm Fe}$ or $T_{\rm{c-m}}$. Furthermore, our models are physically and chemically consistent, rather than assuming that all individual abundances are free parameters, as is the case for the \citet{Benneke2024} and \citet{Holmberg2024} analyses. This means that individual molecular abundances cannot be fine-tuned in order to improve the quality of fit to the spectrum. We also note that the two papers consider different sets of observations: \citet{Benneke2024} fit models to the JWST NIRISS + NIRSpec transmission spectra of the planet, whereas \citet{Holmberg2024} use the HST WFC3 spectrum alongside JWST NIRSpec. Overall, the general agreement between predicted and measured abundances, as well as the low $\chi^2$ values obtained when comparing our models to the JWST spectrum (Figure \ref{fig:spectrum}) indicate that our models are not ruled out by the present JWST observations.

Our models typically find that CO is more abundant in the upper atmosphere than CO$_2$ (see Figure \ref{fig:co_co2}), in agreement with previous studies considering magma-atmosphere interactions at low to moderate redox states \citep{Shorttle2024,Nicholls2024}. Although CO$_2$ has been identified in the atmosphere of TOI-270~d, while CO has not, predicted CO abundances from our models remain below the upper limit suggested by \citet{Benneke2024}. Additional observations to refine the CO abundance measurement may be useful to identify evidence for magma-atmosphere interactions.

\subsection{Detectability of ammonia on sub-Neptunes} \label{subsec:nh3}

Figure \ref{fig:spectrum} shows model transmission spectra for cases in which the atmospheric nitrogen abundance differs by over four orders of magnitude. Despite this, the change in transit depth is relatively small: the difference in $\chi^2/n_{\rm data}$ is just 0.13 between the two models, suggesting they are consistent with each other within $\sim1\sigma$. This highlights that the detection of nitrogen-bearing species such as NH$_3$ on sub-Neptunes will be challenging. Due to its relatively minor impact on the transmission spectrum, the non-detection of NH$_3$ in the atmosphere of this planet from extant transmission spectroscopy does not necessarily imply its absence. Optimal strategies for detecting NH$_3$ may involve stacking multiple transits centered on the 2.9-3.1$\mu$m region, where its absorption is most prominent, as well as acquiring transmission spectra with MIRI/LRS to cover the NH$_3$ feature from 10.3-11$\mu$m \citep[see e.g.,][]{Welbanks2024}. Given the apparent diagnostic power of nitrogen-bearing species for sub-Neptunes, it may be worth investing significant observing time in order to acquire robust detections and abundance measurements.

\subsection{Additional parameters of interest}

The goal of this study was to determine whether magma-atmosphere interactions are sufficient to explain recent JWST observations of the sub-Neptune TOI-270~d. For this reason, we chose to fix the majority of model parameters to nominal values, only varying a small number of parameters of interest. Constructing models with these nominal parameters was sufficient to demonstrate that interactions with a magma ocean can indeed produce atmospheres for TOI-270~d that are consistent with the composition derived by \citet{Benneke2024} from its JWST transmission spectrum. However, we note that it may be of interest in a future study to assess how varying some of these parameters could influence resulting upper atmospheric compositions. For example, \citet{Schlichting2022} noted that the total atmospheric mass fraction of the initial planet (before interacting with the core and mantle) could affect the final atmospheric composition, with larger initial atmospheres leading to higher abundances of H$_2$ and lower abundances of other chemical species. For this work, we chose an initial atmospheric mass fraction of 1.5\%, which is consistent with internal structure models of the planet assuming a H$_2$-dominated atmosphere \citep{VanEylen2021}. In reality, magma-atmosphere interactions likely take place while the planet is still accreting its atmosphere and chemically differentiating internally. Developing a complete model of these interactions would require simultaneously allowing for the influx of new gaseous material as it accretes onto the planet, while allowing the existing atmosphere to react with the core and mantle. Such a model is beyond the scope of this study. We note that \citet{Schlichting2022} found that increasing the total atmospheric mass fraction led to a slight increase in final H$_2$ abundance relative to other chemical species (i.e., lower atmospheric metallicity), but that this effect was less pronounced than that of varying the interface temperatures.

We also assumed core and mantle compositions following \citet{Schlichting2022}. The chosen composition results in an oxygen fugacity relative to the iron-w\"{u}stite buffer $\Delta$IW$=-2.8$, where $\Delta$IW is defined in equation 12 of \citet{Schlichting2022}. A number of studies have demonstrated that varying the oxygen fugacity can alter the atmospheric composition \citep[e.g.,][]{Gaillard2022,Rigby2024,Shorttle2024,Nicholls2024,Nicholls2025}. Although exploring the effect of oxygen fugacity was not required in this study to show that it is possible for our model to explain the observed atmosphere of TOI-270~d, more extensive theoretical work could eventually allow for upper atmospheric composition to be used as a tracer of oxygen fugacity in exoplanetary interiors \citep{Lichtenberg2025}.

\subsection{Model limitations and caveats}

In addition to the need to couple interactions between the core, mantle and atmosphere to evolutionary models as discussed above, \citet{Schlichting2022} note the need to include estimates of the properties of H- and O-rich Fe-bearing metals at the extreme pressures present in the mantles and cores of sub-Neptunes. High-pressure behaviour of the material that comprises planetary interiors remains poorly understood \citep[e.g.,][]{Journaux2020}. Furthermore, it is possible that core-mantle boundary temperatures may exceed the upper limit of 4000~K explored in this work \citep{Young2024}. However, we are unable to produce models at higher temperatures due to a complete lack of relevant laboratory data. We stress the need for further experiments to refine our knowledge of the fundamental quantities underpinning planetary interior and atmospheric models. 

Alongside limitations to the magma-atmosphere interaction model, there are also caveats which apply to the upper atmospheric components of our modeling framework. \texttt{HELIOS} does not yet include the effect of H$_2$O-H$_2$O collision-induced absorption \citep{Anisman2022}, which may be relevant due to the high H$_2$O abundances found by our models. Our modeling framework could also be improved by coupling the radiative transfer and photochemistry in a fully self-consistent manner; however this is not expected to alter major chemical trends \citep{Mukherjee2025}. As noted above, coupling to atmospheric escape models \citep[e.g.,][]{Cherubim2025} will also be an important step towards predicting upper atmospheric composition, particularly at lower atmospheric pressures which may be probed in high-resolution transmission spectroscopy. Additionally, although our framework iterates between a radiative-convective model and an equilibrium chemistry model with rainout condensation, allowing for feedback between the change in chemistry due to condensation and the temperature profile, it does not account for latent heat released by condensation, which can alter the temperature profile \citep{Pierrehumbert2010}.

In addition to H$_2$O, CH$_4$, and CO$_2$, \citet{Benneke2024} find some evidence for the presence of CS$_2$ in the atmosphere of TOI-270~d, albeit with a low detection significance of 2.55$\sigma$. \citet{Lukas2025} also found indications of sulfur-bearing species in this planet's atmosphere. We are unable to assess the likelihood of this species being present as a result of magma-atmosphere interactions, since sulfur-bearing species are not included in the chemical network of the \citet{Schlichting2022} model. Future expansion of the chemical network to include these species, along with nitrogen-bearing species as discussed previously, would enable a more complete picture of the atmospheric composition of atmospheres resulting from magma-atmosphere interactions.

At present, our model does not account for possible demixing between water and hydrogen in the deep atmosphere as proposed by, e.g., \citet{Bailey2021} and \citet{Gupta2024}. However, \citet{Gupta2024} suggest that demixing is unlikely to occur for TOI-270~d, meaning that this limitation should not impact our conclusions regarding this planet.

\subsection{Conclusions}

Interactions between sub-Neptune atmospheres and underlying magma oceans are expected to significantly impact atmospheric composition. We have developed a modeling framework to connect the magma-atmosphere interaction model constructed by \citet{Schlichting2022} to the upper, observable atmosphere. Using this framework, we generated models of the sub-Neptune TOI-270~d in which an initially solar composition atmosphere is allowed to react with the mantle and core. We compared our results to JWST observations of the planet from \citet{Benneke2024}, finding that it is possible to match its measured atmospheric composition, including the abundances of H$_2$O, CH$_4$ and CO$_2$. The model parameters leading to a best-fit solution are $T_{\rm m-a}=3000\,$K, $T_{\rm c-m}=4000\,$K, $x_{\rm Fe}=33\%$, $K_{zz}=10^7\,$cm$^2$s$^{-1}$. We note that $x_{\rm Fe}=33\%$ does not deviate from an Earth-like value, and that many other parameter combinations also provide a reasonable explanation for the planet's measured atmospheric composition. In general, magma-atmosphere interactions appear to enhance atmospheric metallicity while reducing C/O relative to solar values. This indicates that sub-Neptunes with metal-rich atmospheres do not necessarily form by accreting substantial amounts of icy material, and could instead be explained by the accretion of hydrogen-dominated nebular gas onto a molten rocky body.

This work represents an early step towards understanding just one process controlling how sub-Neptunes formed and evolved to their present-day states. Substantial further work is required to understand the relative contributions affecting sub-Neptune atmospheric composition and evolution. However, with a combination of new modeling efforts and high-quality spectra now available from JWST, the promise of uncovering the nature of these mysterious worlds has never been greater.

\begin{acknowledgments}

The authors thank the anonymous referee for their careful review of this manuscript. This research was supported by the AEThER program, funded in part by the Alfred P.\ Sloan Foundation under grant \#G202114194, as well as by NASA ADAP 80NSSC19K1014. The authors thank Anat Shahar, the PI of the AEThER collaboration, as well as other collaboration members for fruitful discussion that helped to shape many of the ideas behind this work. Matthew Nixon thanks the Heising-Simons Foundation for their funding through the 51 Pegasi b Postdoctoral Fellowship. This research has made use of the NASA Astrophysics Data System and the NASA Exoplanet Archive. Thermodynamic data required for the magma-atmosphere interaction model was provided in part by the National Institute of Standards and Technology Standard Reference Data Program.

\end{acknowledgments}

\software{\textsc{NumPy} \citep{Harris2020},  
          \textsc{SciPy} \citep{Virtanen2020}, 
          \textsc{Matplotlib} \citep{Hunter2007}
          }

\vspace{5mm}

\bibliography{references}

@ARTICLE{Tang2025,
       author = {{Tang}, Yao and {Fortney}, Jonathan J. and {Nimmo}, Francis and {Thorngren}, Daniel and {Ohno}, Kazumasa and {Murray-Clay}, Ruth},
        title = "{Reassessing Sub-Neptune Structure, Radii, and Thermal Evolution}",
      journal = {\apj},
         year = 2025,
        month = aug,
       volume = {989},
       number = {1},
          eid = {28},
        pages = {28},
          doi = {10.3847/1538-4357/ade7ff},
archivePrefix = {arXiv},
       eprint = {2410.21584},
 primaryClass = {astro-ph.EP},
       adsurl = {https://ui.adsabs.harvard.edu/abs/2025ApJ...989...28T}
}

@ARTICLE{Wogan2025,
       author = {{Wogan}, Nicholas F. and {Batalha}, Natasha E. and {Zahnle}, Kevin and {Krissansen-Totton}, Joshua and {Catling}, David C. and {Wolf}, Eric T. and {Robinson}, Tyler D. and {Meadows}, Victoria and {Arney}, Giada and {Domagal-Goldman}, Shawn},
        title = "{The Open-Source Photochem Code: A General Chemical and Climate Model for Interpreting (Exo)Planet Observations}",
      journal = {arXiv e-prints},
         year = 2025,
        month = sep,
          eid = {arXiv:2509.25578},
        pages = {arXiv:2509.25578},
          doi = {10.48550/arXiv.2509.25578},
archivePrefix = {arXiv},
       eprint = {2509.25578},
 primaryClass = {astro-ph.EP},
       adsurl = {https://ui.adsabs.harvard.edu/abs/2025arXiv250925578W}
}

@ARTICLE{Breza2025,
       author = {{Breza}, Bodie and {Nixon}, Matthew C. and {Kempton}, Eliza M. -R.},
        title = "{Not All Sub-Neptune Exoplanets Have Magma Oceans}",
      journal = {arXiv e-prints},
         year = 2025,
        month = sep,
          eid = {arXiv:2509.20429},
        pages = {arXiv:2509.20429},
archivePrefix = {arXiv},
       eprint = {2509.20429},
 primaryClass = {astro-ph.EP},
       adsurl = {https://ui.adsabs.harvard.edu/abs/2025arXiv250920429B}
}

@BOOK{Pierrehumbert2010,
       author = {{Pierrehumbert}, Raymond T.},
        title = "{Principles of Planetary Climate}",
         year = 2010,
    publisher = {Cambridge University Press},
         city = {Cambridge},
       adsurl = {https://ui.adsabs.harvard.edu/abs/2010ppc..book.....P}
}

@ARTICLE{Hansen2008,
       author = {{Hansen}, Brad M.~S.},
        title = "{On the Absorption and Redistribution of Energy in Irradiated Planets}",
      journal = {\apjs},
         year = 2008,
        month = dec,
       volume = {179},
       number = {2},
        pages = {484-508},
          doi = {10.1086/591964},
archivePrefix = {arXiv},
       eprint = {0801.2972},
 primaryClass = {astro-ph},
       adsurl = {https://ui.adsabs.harvard.edu/abs/2008ApJS..179..484H}
}

@ARTICLE{Bailey2021,
       author = {{Bailey}, Elizabeth and {Stevenson}, David J.},
        title = "{Thermodynamically Governed Interior Models of Uranus and Neptune}",
      journal = {\psj},
         year = 2021,
        month = apr,
       volume = {2},
       number = {2},
          eid = {64},
        pages = {64},
          doi = {10.3847/PSJ/abd1e0},
archivePrefix = {arXiv},
       eprint = {2012.04166},
 primaryClass = {astro-ph.EP},
       adsurl = {https://ui.adsabs.harvard.edu/abs/2021PSJ.....2...64B}
}

@book{Barin1995,
  author    = {{Barin}, I.},
  title     = {Thermochemical Data of Pure Substances},
  publisher = {VCH},
  city      = {Weinheim},
  year      = {1995},
}

@ARTICLE{Gail2013,
       author = {{Gail}, H. -P. and {Wetzel}, S. and {Pucci}, A. and {Tamanai}, A.},
        title = "{Seed particle formation for silicate dust condensation by SiO nucleation}",
      journal = {\aap},
         year = 2013,
        month = jul,
       volume = {555},
          eid = {A119},
        pages = {A119},
          doi = {10.1051/0004-6361/201321807},
archivePrefix = {arXiv},
       eprint = {1305.2879},
 primaryClass = {astro-ph.SR},
       adsurl = {https://ui.adsabs.harvard.edu/abs/2013A&A...555A.119G}
}

@ARTICLE{Haar1978,
       author = {{Haar}, Lester and {Gallagher}, John S.},
        title = "{Thermodynamic properties of ammonia}",
      journal = {Journal of Physical and Chemical Reference Data},
         year = 1978,
        month = jul,
       volume = {7},
       number = {3},
        pages = {635-792},
          doi = {10.1063/1.555579},
       adsurl = {https://ui.adsabs.harvard.edu/abs/1978JPCRD...7..635H}
}

@book{Lide2009,
  author    = {{Lide}, D},
  title     = {CRC Handbook of Chemistry and Physics, 90th edition},
  publisher = {CRC Press},
  city      = {Boca Raton},
  year      = {2009},
}

@book{Dykyj2001,
  author    = {{Dykyj}, J. and {Svoboda}, J. and {Wilhoit}, R. C. and {Hall}, K. R.},
  title     = {Landolt-B\"{o}rnstein – Group IV Physical Chemistry, Volume 20C, Vapor Pressure and Antoine Constants for Nitrogen Containing Organic Compounds},
  publisher = {Springer-Verlag},
  city      = {Berlin},
  year      = {2001},
}

@ARTICLE{Murphy2005,
       author = {{Murphy}, D.~M. and {Koop}, T.},
        title = "{Review of the vapour pressures of ice and supercooled water for atmospheric applications}",
      journal = {Quarterly Journal of the Royal Meteorological Society},
         year = 2005,
        month = apr,
       volume = {131},
       number = {608},
        pages = {1539-1565},
          doi = {10.1256/qj.04.94},
       adsurl = {https://ui.adsabs.harvard.edu/abs/2005QJRMS.131.1539M}
}

@ARTICLE{Sharp1990,
       author = {{Sharp}, C.~M. and {Huebner}, W.~F.},
        title = "{Molecular Equilibrium with Condensation}",
      journal = {\apjs},
         year = 1990,
        month = feb,
       volume = {72},
        pages = {417},
          doi = {10.1086/191422},
       adsurl = {https://ui.adsabs.harvard.edu/abs/1990ApJS...72..417S}
}

@book{Wagner2008,
  author    = {{Wagner}, W. and {Kretzschmar}, H.-J.},
  title     = {International Steam Tables: Properties of Water and Steam Based on the Industrial Formulation IAPWS-IF97},
  publisher = {Springer-Verlag},
  city      = {Berlin},
  year      = {2008},
}

@book{Yaws1999,
  author    = {Carl L. Yaws},
  title     = {Chemical Properties Handbook: Physical, Thermodynamic, Environmental, Transport, Safety and Health Related Properties for Organic and Inorganic Chemicals},
  publisher = {McGraw-Hill},
  city      = {New York},
  year      = {1999},
  isbn      = {9780070734012}
}

@ARTICLE{Goodwin1985,
       author = {{Goodwin}, Robert D.},
        title = "{Carbon Monoxide Thermophysical Properties from 68 to 1000 K at Pressures to 100 MPa}",
      journal = {Journal of Physical and Chemical Reference Data},
         year = 1985,
        month = oct,
       volume = {14},
       number = {4},
        pages = {849-932},
          doi = {10.1063/1.555742},
       adsurl = {https://ui.adsabs.harvard.edu/abs/1985JPCRD..14..849G}
}

@article{Prydz1972,
  author  = {Prydz, R. and Goodwin, R. D.},
  title   = {Experimental melting and vapor pressures of methane},
  journal = {The Journal of Chemical Thermodynamics},
  year    = {1972},
  volume  = {4},
  month   = jan,
  number  = {1},
  pages   = {127–33},
  doi     = {10.1016/S0021-9614(72)80016-8}
}

@ARTICLE{Moses1992,
       author = {{Moses}, Julianne I. and {Allen}, Mark and {Yung}, Yuk L.},
        title = "{Hydrocarbon nucleation and aerosol formation in Neptune's atmosphere}",
      journal = {\icarus},
         year = 1992,
        month = oct,
       volume = {99},
       number = {2},
        pages = {318-346},
          doi = {10.1016/0019-1035(92)90149-2},
       adsurl = {https://ui.adsabs.harvard.edu/abs/1992Icar...99..318M}
}

@book{Chase1998,
  author = {M Chase},
  title = {NIST-JANAF Thermochemical Tables, 4th Edition},
  year = {1998},
  month = {1998-08-01},
  publisher = {American Institute of Physics, -1},
  language = {en},
}

@ARTICLE{Pan1991,
       author = {{Pan}, Vivian and {Holloway}, John R. and {Hervig}, Richard L.},
        title = "{The pressure and temperature dependence of carbon dioxide solubility in tholeiitic basalt melts}",
      journal = {\gca},
         year = 1991,
        month = jun,
       volume = {55},
       number = {6},
        pages = {1587-1595},
          doi = {10.1016/0016-7037(91)90130-W},
       adsurl = {https://ui.adsabs.harvard.edu/abs/1991GeCoA..55.1587P}
}

@ARTICLE{Hirschmann2016,
       author = {{Hirschmann}, Marc M.},
        title = "{Constraints on the early delivery and fractionation of Earth's major volatiles from C/H, C/N, and C/S ratios}",
      journal = {American Mineralogist},
         year = 2016,
        month = mar,
       volume = {101},
       number = {3},
        pages = {540-553},
          doi = {10.2138/am-2016-5452},
       adsurl = {https://ui.adsabs.harvard.edu/abs/2016AmMin.101..540H}
}

@ARTICLE{Moore1998,
       author = {{Moore}, Gordon and {Vennemann}, Torsten and {Carmichael}, I.~S.~E.},
        title = "{An empirical model for the solubility of H2O in magmas to 3 kilobars}",
      journal = {American Mineralogist},
         year = 1998,
        month = feb,
       volume = {83},
       number = {1-2},
        pages = {36-42},
          doi = {10.2138/am-1998-1-203},
       adsurl = {https://ui.adsabs.harvard.edu/abs/1998AmMin..83...36M}
}

@ARTICLE{Okuchi1997,
       author = {{Okuchi}, Takuo},
        title = "{Hydrogen Partitioning into Molten Iron at High Pressure: Implications for Earth's Core}",
      journal = {Science},
         year = 1997,
        month = dec,
       volume = {278},
        pages = {1781},
          doi = {10.1126/science.278.5344.1781},
       adsurl = {https://ui.adsabs.harvard.edu/abs/1997Sci...278.1781O}
}

@ARTICLE{Hirschmann2012,
       author = {{Hirschmann}, M.~M. and {Withers}, A.~C. and {Ardia}, P. and {Foley}, N.~T.},
        title = "{Solubility of molecular hydrogen in silicate melts and consequences for volatile evolution of terrestrial planets}",
      journal = {Earth and Planetary Science Letters},
         year = 2012,
        month = sep,
       volume = {345},
        pages = {38-48},
          doi = {10.1016/j.epsl.2012.06.031},
       adsurl = {https://ui.adsabs.harvard.edu/abs/2012E&PSL.345...38H}
}

@ARTICLE{Badro2015,
       author = {{Badro}, James and {Brodholt}, John P. and {Piet}, H{\'e}l{\`e}ne and {Siebert}, Julien and {Ryerson}, Frederick J.},
        title = "{Core formation and core composition from coupled geochemical and geophysical constraints}",
      journal = {Proceedings of the National Academy of Science},
         year = 2015,
        month = oct,
       volume = {112},
       number = {40},
        pages = {12310-12314},
          doi = {10.1073/pnas.1505672112},
       adsurl = {https://ui.adsabs.harvard.edu/abs/2015PNAS..11212310B}
}

@ARTICLE{Fegley1987,
       author = {{Fegley}, B. and {Cameron}, A.~G.~W.},
        title = "{A vaporization model for iron/silicate fractionation in the Mercury protoplanet}",
      journal = {Earth and Planetary Science Letters},
         year = 1987,
        month = apr,
       volume = {82},
       number = {3-4},
        pages = {207-222},
          doi = {10.1016/0012-821X(87)90196-8},
       adsurl = {https://ui.adsabs.harvard.edu/abs/1987E&PSL..82..207F}
}

@ARTICLE{Beatty2024,
       author = {{Beatty}, Thomas G. and {Welbanks}, Luis and {Schlawin}, Everett and {Bell}, Taylor J. and {Line}, Michael R. and {Murphy}, Matthew and {Edelman}, Isaac and {Greene}, Thomas P. and {Fortney}, Jonathan J. and {Henry}, Gregory W. and {Mukherjee}, Sagnick and {Ohno}, Kazumasa and {Parmentier}, Vivien and {Rauscher}, Emily and {Wiser}, Lindsey S. and {Arnold}, Kenneth E.},
        title = "{Sulfur Dioxide and Other Molecular Species in the Atmosphere of the Sub-Neptune GJ 3470 b}",
      journal = {\apjl},
         year = 2024,
        month = jul,
       volume = {970},
       number = {1},
          eid = {L10},
        pages = {L10},
          doi = {10.3847/2041-8213/ad55e9},
archivePrefix = {arXiv},
       eprint = {2406.04450},
 primaryClass = {astro-ph.EP},
       adsurl = {https://ui.adsabs.harvard.edu/abs/2024ApJ...970L..10B}
}

@ARTICLE{Pineda2021,
       author = {{Pineda}, J. Sebastian and {Youngblood}, Allison and {France}, Kevin},
        title = "{The M-dwarf Ultraviolet Spectroscopic Sample. I. Determining Stellar Parameters for Field Stars}",
      journal = {\apj},
         year = 2021,
        month = sep,
       volume = {918},
       number = {1},
          eid = {40},
        pages = {40},
          doi = {10.3847/1538-4357/ac0aea},
archivePrefix = {arXiv},
       eprint = {2106.07656},
 primaryClass = {astro-ph.SR},
       adsurl = {https://ui.adsabs.harvard.edu/abs/2021ApJ...918...40P}
}

@ARTICLE{Zahnle2016,
       author = {{Zahnle}, K. and {Marley}, M.~S. and {Morley}, C.~V. and {Moses}, J.~I.},
        title = "{Photolytic Hazes in the Atmosphere of 51 Eri b}",
      journal = {\apj},
         year = 2016,
        month = jun,
       volume = {824},
       number = {2},
          eid = {137},
        pages = {137},
          doi = {10.3847/0004-637X/824/2/137},
archivePrefix = {arXiv},
       eprint = {1604.07388},
 primaryClass = {astro-ph.EP},
       adsurl = {https://ui.adsabs.harvard.edu/abs/2016ApJ...824..137Z}
}

@software{photochem,
       author = {{Wogan}, Nick},
        title = "{Nicholaswogan/photochem: photochem v0.6.2}",
         year = 2024,
        month = nov,
          eid = {10.5281/zenodo.14032108},
          doi = {10.5281/zenodo.14032108},
      version = {v0.6.2},
    publisher = {Zenodo},
       adsurl = {https://ui.adsabs.harvard.edu/abs/2024zndo..14032108W}
}

@ARTICLE{Mukherjee2025,
       author = {{Mukherjee}, Sagnick and {Fortney}, Jonathan J. and {Wogan}, Nicholas F. and {Sing}, David K. and {Ohno}, Kazumasa},
        title = "{Effects of Planetary Parameters on Disequilibrium Chemistry in Irradiated Planetary Atmospheres: From Gas Giants to Sub-Neptunes}",
      journal = {\apj},
         year = 2025,
        month = jun,
       volume = {985},
       number = {2},
          eid = {209},
        pages = {209},
          doi = {10.3847/1538-4357/adc7b3},
archivePrefix = {arXiv},
       eprint = {2410.17169},
 primaryClass = {astro-ph.EP},
       adsurl = {https://ui.adsabs.harvard.edu/abs/2025ApJ...985..209M}
}

@ARTICLE{Glein2025,
       author = {{Glein}, Christopher R. and {Yu}, Xinting and {Luu}, Cindy N.},
        title = "{Deciphering Sub-Neptune Atmospheres: New Insights from Geochemical Models of TOI-270 d}",
      journal = {\apj},
         year = 2025,
        month = jun,
       volume = {985},
       number = {2},
          eid = {187},
        pages = {187},
          doi = {10.3847/1538-4357/adced4},
archivePrefix = {arXiv},
       eprint = {2504.09752},
 primaryClass = {astro-ph.EP},
       adsurl = {https://ui.adsabs.harvard.edu/abs/2025ApJ...985..187G}
}

@ARTICLE{Lukas2025,
       author = {{Felix}, Lukas and {Kitzmann}, Daniel and {Demory}, Brice-Olivier and {Mordasini}, Christoph},
        title = "{Evidence for sulfur chemistry in the atmosphere of the warm sub-Neptune TOI-270 d}",
      journal = {arXiv e-prints},
         year = 2025,
        month = apr,
          eid = {arXiv:2504.13039},
        pages = {arXiv:2504.13039},
          doi = {10.48550/arXiv.2504.13039},
archivePrefix = {arXiv},
       eprint = {2504.13039},
 primaryClass = {astro-ph.EP},
       adsurl = {https://ui.adsabs.harvard.edu/abs/2025arXiv250413039F}
}

@ARTICLE{Moses2005,
       author = {{Moses}, J.~I. and {Fouchet}, T. and {B{\'e}zard}, B. and {Gladstone}, G.~R. and {Lellouch}, E. and {Feuchtgruber}, H.},
        title = "{Photochemistry and diffusion in Jupiter's stratosphere: Constraints from ISO observations and comparisons with other giant planets}",
      journal = {Journal of Geophysical Research (Planets)},
         year = 2005,
        month = aug,
       volume = {110},
       number = {E8},
          eid = {E08001},
        pages = {E08001},
          doi = {10.1029/2005JE002411},
       adsurl = {https://ui.adsabs.harvard.edu/abs/2005JGRE..110.8001M}
}

@ARTICLE{Lichtenberg2022,
       author = {{Lichtenberg}, Tim and {Clement}, Matthew S.},
        title = "{Reduced Late Bombardment on Rocky Exoplanets around M Dwarfs}",
      journal = {\apjl},
         year = 2022,
        month = oct,
       volume = {938},
       number = {1},
          eid = {L3},
        pages = {L3},
          doi = {10.3847/2041-8213/ac9521},
archivePrefix = {arXiv},
       eprint = {2209.14037},
 primaryClass = {astro-ph.EP},
       adsurl = {https://ui.adsabs.harvard.edu/abs/2022ApJ...938L...3L}
}

@ARTICLE{Anisman2022,
       author = {{Anisman}, Lara O. and {Chubb}, Katy L. and {Changeat}, Quentin and {Edwards}, Billy and {Yurchenko}, Sergei N. and {Tennyson}, Jonathan and {Tinetti}, Giovanna},
        title = "{Cross-sections for heavy atmospheres: H$_{2}$O self-broadening}",
      journal = {\jqsrt},
         year = 2022,
        month = jun,
       volume = {283},
          eid = {108146},
        pages = {108146},
          doi = {10.1016/j.jqsrt.2022.108146},
archivePrefix = {arXiv},
       eprint = {2203.02335},
 primaryClass = {astro-ph.EP},
       adsurl = {https://ui.adsabs.harvard.edu/abs/2022JQSRT.28308146A}
}

@ARTICLE{Amundsen2017,
       author = {{Amundsen}, David S. and {Tremblin}, Pascal and {Manners}, James and {Baraffe}, Isabelle and {Mayne}, Nathan J.},
        title = "{Treatment of overlapping gaseous absorption with the correlated-k method in hot Jupiter and brown dwarf atmosphere models}",
      journal = {\aap},
         year = 2017,
        month = feb,
       volume = {598},
          eid = {A97},
        pages = {A97},
          doi = {10.1051/0004-6361/201629322},
archivePrefix = {arXiv},
       eprint = {1610.01389},
 primaryClass = {astro-ph.EP},
       adsurl = {https://ui.adsabs.harvard.edu/abs/2017A&A...598A..97A}
}

@ARTICLE{Wogan2024,
       author = {{Wogan}, Nicholas F. and {Batalha}, Natasha E. and {Zahnle}, Kevin J. and {Krissansen-Totton}, Joshua and {Tsai}, Shang-Min and {Hu}, Renyu},
        title = "{JWST Observations of K2-18b Can Be Explained by a Gas-rich Mini-Neptune with No Habitable Surface}",
      journal = {\apjl},
         year = 2024,
        month = mar,
       volume = {963},
       number = {1},
          eid = {L7},
        pages = {L7},
          doi = {10.3847/2041-8213/ad2616},
archivePrefix = {arXiv},
       eprint = {2401.11082},
 primaryClass = {astro-ph.EP},
       adsurl = {https://ui.adsabs.harvard.edu/abs/2024ApJ...963L...7W}
}

@ARTICLE{Grimm2021,
       author = {{Grimm}, Simon L. and {Malik}, Matej and {Kitzmann}, Daniel and {Guzm{\'a}n-Mesa}, Andrea and {Hoeijmakers}, H. Jens and {Fisher}, Chloe and {Mendon{\c{c}}a}, Jo{\~a}o M. and {Yurchenko}, Sergey N. and {Tennyson}, Jonathan and {Alesina}, Fabien and {Buchschacher}, Nicolas and {Burnier}, Julien and {Segransan}, Damien and {Kurucz}, Robert L. and {Heng}, Kevin},
        title = "{HELIOS-K 2.0 Opacity Calculator and Open-source Opacity Database for Exoplanetary Atmospheres}",
      journal = {\apjs},
         year = 2021,
        month = mar,
       volume = {253},
       number = {1},
          eid = {30},
        pages = {30},
          doi = {10.3847/1538-4365/abd773},
archivePrefix = {arXiv},
       eprint = {2101.02005},
 primaryClass = {astro-ph.EP},
       adsurl = {https://ui.adsabs.harvard.edu/abs/2021ApJS..253...30G}
}

@ARTICLE{Werlen2025,
       author = {{Werlen}, Aaron and {Dorn}, Caroline and {Schlichting}, Hilke E. and {Grimm}, Simon L. and {Young}, Edward D.},
        title = "{Atmospheric C/O Ratios of Sub-Neptunes with Magma Oceans: Homemade rather than Inherited}",
      journal = {arXiv e-prints},
         year = 2025,
        month = apr,
          eid = {arXiv:2504.20450},
        pages = {arXiv:2504.20450},
          doi = {10.48550/arXiv.2504.20450},
archivePrefix = {arXiv},
       eprint = {2504.20450},
 primaryClass = {astro-ph.EP},
       adsurl = {https://ui.adsabs.harvard.edu/abs/2025arXiv250420450W}
}

@ARTICLE{Cherubim2025,
       author = {{Cherubim}, Collin and {Wordsworth}, Robin and {Bower}, Dan J. and {Sossi}, Paolo A. and {Adams}, Danica and {Hu}, Renyu},
        title = "{An Oxidation Gradient Straddling the Small Planet Radius Valley}",
      journal = {\apj},
         year = 2025,
        month = apr,
       volume = {983},
       number = {2},
          eid = {97},
        pages = {97},
          doi = {10.3847/1538-4357/adbca9},
archivePrefix = {arXiv},
       eprint = {2503.05055},
 primaryClass = {astro-ph.EP},
       adsurl = {https://ui.adsabs.harvard.edu/abs/2025ApJ...983...97C}
}

@ARTICLE{Burrows1999,
       author = {{Burrows}, Adam and {Sharp}, C.~M.},
        title = "{Chemical Equilibrium Abundances in Brown Dwarf and Extrasolar Giant Planet Atmospheres}",
      journal = {\apj},
         year = 1999,
        month = feb,
       volume = {512},
       number = {2},
        pages = {843-863},
          doi = {10.1086/306811},
archivePrefix = {arXiv},
       eprint = {astro-ph/9807055},
 primaryClass = {astro-ph},
       adsurl = {https://ui.adsabs.harvard.edu/abs/1999ApJ...512..843B}
}

@ARTICLE{Young2024,
       author = {{Young}, Edward D. and {Stixrude}, Lars and {Rogers}, James G. and {Schlichting}, Hilke E. and {Marcum}, Sarah P.},
        title = "{Phase Equilibria of Sub-Neptunes and Super-Earths}",
      journal = {\psj},
         year = 2024,
        month = dec,
       volume = {5},
       number = {12},
          eid = {268},
        pages = {268},
          doi = {10.3847/PSJ/ad8c40},
archivePrefix = {arXiv},
       eprint = {2408.11321},
 primaryClass = {astro-ph.EP},
       adsurl = {https://ui.adsabs.harvard.edu/abs/2024PSJ.....5..268Y}
}

@ARTICLE{Heng2025,
       author = {{Heng}, Kevin and {Owen}, James E. and {Tian}, Meng},
        title = "{The Gradient of Mean Molecular Weight Across the Radius Valley}",
      journal = {arXiv e-prints},
         year = 2025,
        month = apr,
          eid = {arXiv:2504.02499},
        pages = {arXiv:2504.02499},
          doi = {10.48550/arXiv.2504.02499},
archivePrefix = {arXiv},
       eprint = {2504.02499},
 primaryClass = {astro-ph.EP},
       adsurl = {https://ui.adsabs.harvard.edu/abs/2025arXiv250402499H}
}

@ARTICLE{Rogers2023,
       author = {{Rogers}, James G. and {Schlichting}, Hilke E. and {Owen}, James E.},
        title = "{Conclusive Evidence for a Population of Water Worlds around M Dwarfs Remains Elusive}",
      journal = {\apjl},
         year = 2023,
        month = apr,
       volume = {947},
       number = {1},
          eid = {L19},
        pages = {L19},
          doi = {10.3847/2041-8213/acc86f},
archivePrefix = {arXiv},
       eprint = {2301.04321},
 primaryClass = {astro-ph.EP},
       adsurl = {https://ui.adsabs.harvard.edu/abs/2023ApJ...947L..19R}
}

@ARTICLE{Lichtenberg2021,
       author = {{Lichtenberg}, Tim},
        title = "{Redox Hysteresis of Super-Earth Exoplanets from Magma Ocean Circulation}",
      journal = {\apjl},
         year = 2021,
        month = jun,
       volume = {914},
       number = {1},
          eid = {L4},
        pages = {L4},
          doi = {10.3847/2041-8213/ac0146},
archivePrefix = {arXiv},
       eprint = {2105.11208},
 primaryClass = {astro-ph.EP},
       adsurl = {https://ui.adsabs.harvard.edu/abs/2021ApJ...914L...4L}
}

@ARTICLE{Kite2020,
       author = {{Kite}, Edwin S. and {Fegley}, Jr., Bruce and {Schaefer}, Laura and {Ford}, Eric B.},
        title = "{Atmosphere Origins for Exoplanet Sub-Neptunes}",
      journal = {\apj},
         year = 2020,
        month = mar,
       volume = {891},
       number = {2},
          eid = {111},
        pages = {111},
          doi = {10.3847/1538-4357/ab6ffb},
archivePrefix = {arXiv},
       eprint = {2001.09269},
 primaryClass = {astro-ph.EP},
       adsurl = {https://ui.adsabs.harvard.edu/abs/2020ApJ...891..111K}
}

@ARTICLE{Lichtenberg2025,
       author = {{Lichtenberg}, Tim and {Miguel}, Yamila},
        title = "{Super-Earths and Earth-like Exoplanets}",
      journal = {Treatise on Geochemistry},
         year = 2025,
        month = jan,
       volume = {7},
        pages = {51-112},
          doi = {10.1016/B978-0-323-99762-1.00122-4},
archivePrefix = {arXiv},
       eprint = {2405.04057},
 primaryClass = {astro-ph.EP},
       adsurl = {https://ui.adsabs.harvard.edu/abs/2025TrGeo...7...51L}
}

@ARTICLE{Tsai2021,
       author = {{Tsai}, Shang-Min and {Innes}, Hamish and {Lichtenberg}, Tim and {Taylor}, Jake and {Malik}, Matej and {Chubb}, Katy and {Pierrehumbert}, Raymond},
        title = "{Inferring Shallow Surfaces on Sub-Neptune Exoplanets with JWST}",
      journal = {\apjl},
         year = 2021,
        month = dec,
       volume = {922},
       number = {2},
          eid = {L27},
        pages = {L27},
          doi = {10.3847/2041-8213/ac399a},
archivePrefix = {arXiv},
       eprint = {2111.06429},
 primaryClass = {astro-ph.EP},
       adsurl = {https://ui.adsabs.harvard.edu/abs/2021ApJ...922L..27T}
}

@ARTICLE{Yu2021,
       author = {{Yu}, Xinting and {Moses}, Julianne I. and {Fortney}, Jonathan J. and {Zhang}, Xi},
        title = "{How to Identify Exoplanet Surfaces Using Atmospheric Trace Species in Hydrogen-dominated Atmospheres}",
      journal = {\apj},
         year = 2021,
        month = jun,
       volume = {914},
       number = {1},
          eid = {38},
        pages = {38},
          doi = {10.3847/1538-4357/abfdc7},
archivePrefix = {arXiv},
       eprint = {2104.09843},
 primaryClass = {astro-ph.EP},
       adsurl = {https://ui.adsabs.harvard.edu/abs/2021ApJ...914...38Y}
}

@ARTICLE{Hu2021,
       author = {{Hu}, Renyu and {Damiano}, Mario and {Scheucher}, Markus and {Kite}, Edwin and {Seager}, Sara and {Rauer}, Heike},
        title = "{Unveiling Shrouded Oceans on Temperate sub-Neptunes via Transit Signatures of Solubility Equilibria versus Gas Thermochemistry}",
      journal = {\apjl},
         year = 2021,
        month = nov,
       volume = {921},
       number = {1},
          eid = {L8},
        pages = {L8},
          doi = {10.3847/2041-8213/ac1f92},
archivePrefix = {arXiv},
       eprint = {2108.04745},
 primaryClass = {astro-ph.EP},
       adsurl = {https://ui.adsabs.harvard.edu/abs/2021ApJ...921L...8H}
}

@ARTICLE{Nicholls2024,
       author = {{Nicholls}, Harrison and {Lichtenberg}, Tim and {Bower}, Dan J. and {Pierrehumbert}, Raymond},
        title = "{Magma Ocean Evolution at Arbitrary Redox State}",
      journal = {Journal of Geophysical Research (Planets)},
         year = 2024,
        month = dec,
       volume = {129},
       number = {12},
        pages = {2024JE008576},
          doi = {10.1029/2024JE008576},
archivePrefix = {arXiv},
       eprint = {2411.19137},
 primaryClass = {astro-ph.EP},
       adsurl = {https://ui.adsabs.harvard.edu/abs/2024JGRE..12908576N}
}

@ARTICLE{Nicholls2025,
       author = {{Nicholls}, Harrison and {Pierrehumbert}, Raymond T. and {Lichtenberg}, Tim and {Soucasse}, Laurent and {Smeets}, Stef},
        title = "{Convective shutdown in the atmospheres of lava worlds}",
      journal = {\mnras},
         year = 2025,
        month = jan,
       volume = {536},
       number = {3},
        pages = {2957-2971},
          doi = {10.1093/mnras/stae2772},
archivePrefix = {arXiv},
       eprint = {2412.11987},
 primaryClass = {astro-ph.EP},
       adsurl = {https://ui.adsabs.harvard.edu/abs/2025MNRAS.536.2957N}
}

@ARTICLE{France2016,
       author = {{France}, Kevin and {Loyd}, R.~O. Parke and {Youngblood}, Allison and {Brown}, Alexander and {Schneider}, P. Christian and {Hawley}, Suzanne L. and {Froning}, Cynthia S. and {Linsky}, Jeffrey L. and {Roberge}, Aki and {Buccino}, Andrea P. and {Davenport}, James R.~A. and {Fontenla}, Juan M. and {Kaltenegger}, Lisa and {Kowalski}, Adam F. and {Mauas}, Pablo J.~D. and {Miguel}, Yamila and {Redfield}, Seth and {Rugheimer}, Sarah and {Tian}, Feng and {Vieytes}, Mariela C. and {Walkowicz}, Lucianne M. and {Weisenburger}, Kolby L.},
        title = "{The MUSCLES Treasury Survey. I. Motivation and Overview}",
      journal = {\apj},
         year = 2016,
        month = apr,
       volume = {820},
       number = {2},
          eid = {89},
        pages = {89},
          doi = {10.3847/0004-637X/820/2/89},
archivePrefix = {arXiv},
       eprint = {1602.09142},
 primaryClass = {astro-ph.SR},
       adsurl = {https://ui.adsabs.harvard.edu/abs/2016ApJ...820...89F}
}

@ARTICLE{Youngblood2016,
       author = {{Youngblood}, Allison and {France}, Kevin and {Loyd}, R.~O. Parke and {Linsky}, Jeffrey L. and {Redfield}, Seth and {Schneider}, P. Christian and {Wood}, Brian E. and {Brown}, Alexander and {Froning}, Cynthia and {Miguel}, Yamila and {Rugheimer}, Sarah and {Walkowicz}, Lucianne},
        title = "{The MUSCLES Treasury Survey. II. Intrinsic LY{\ensuremath{\alpha}} and Extreme Ultraviolet Spectra of K and M Dwarfs with Exoplanets*}",
      journal = {\apj},
         year = 2016,
        month = jun,
       volume = {824},
       number = {2},
          eid = {101},
        pages = {101},
          doi = {10.3847/0004-637X/824/2/101},
archivePrefix = {arXiv},
       eprint = {1604.01032},
 primaryClass = {astro-ph.SR},
       adsurl = {https://ui.adsabs.harvard.edu/abs/2016ApJ...824..101Y}
}

@ARTICLE{Lloyd2016,
       author = {{Loyd}, R.~O.~P. and {France}, Kevin and {Youngblood}, Allison and {Schneider}, Christian and {Brown}, Alexander and {Hu}, Renyu and {Linsky}, Jeffrey and {Froning}, Cynthia S. and {Redfield}, Seth and {Rugheimer}, Sarah and {Tian}, Feng},
        title = "{The MUSCLES Treasury Survey. III. X-Ray to Infrared Spectra of 11 M and K Stars Hosting Planets}",
      journal = {\apj},
         year = 2016,
        month = jun,
       volume = {824},
       number = {2},
          eid = {102},
        pages = {102},
          doi = {10.3847/0004-637X/824/2/102},
archivePrefix = {arXiv},
       eprint = {1604.04776},
 primaryClass = {astro-ph.SR},
       adsurl = {https://ui.adsabs.harvard.edu/abs/2016ApJ...824..102L}
}

@ARTICLE{Gaillard2022,
       author = {{Gaillard}, Fabrice and {Bernadou}, Fabien and {Roskosz}, Mathieu and {Bouhifd}, Mohamed Ali and {Marrocchi}, Yves and {Iacono-Marziano}, Giada and {Moreira}, Manuel and {Scaillet}, Bruno and {Rogerie}, Gregory},
        title = "{Redox controls during magma ocean degassing}",
      journal = {Earth and Planetary Science Letters},
         year = 2022,
        month = jan,
       volume = {577},
          eid = {117255},
        pages = {117255},
          doi = {10.1016/j.epsl.2021.117255},
       adsurl = {https://ui.adsabs.harvard.edu/abs/2022E&PSL.57717255G}
}

@ARTICLE{Schulik2023,
       author = {{Schulik}, Matth{\"a}us and {Booth}, Richard A.},
        title = "{AIOLOS - A multipurpose 1D hydrodynamics code for planetary atmospheres}",
      journal = {\mnras},
         year = 2023,
        month = jul,
       volume = {523},
       number = {1},
        pages = {286-304},
          doi = {10.1093/mnras/stad1251},
archivePrefix = {arXiv},
       eprint = {2207.07144},
 primaryClass = {astro-ph.EP},
       adsurl = {https://ui.adsabs.harvard.edu/abs/2023MNRAS.523..286S}
}

@ARTICLE{Andrae2010,
       author = {{Andrae}, Rene and {Schulze-Hartung}, Tim and {Melchior}, Peter},
        title = "{Dos and don'ts of reduced chi-squared}",
      journal = {arXiv e-prints},
         year = 2010,
        month = dec,
          eid = {arXiv:1012.3754},
        pages = {arXiv:1012.3754},
          doi = {10.48550/arXiv.1012.3754},
archivePrefix = {arXiv},
       eprint = {1012.3754},
 primaryClass = {astro-ph.IM},
       adsurl = {https://ui.adsabs.harvard.edu/abs/2010arXiv1012.3754A}
}

@ARTICLE{Gupta2019,
       author = {{Gupta}, Akash and {Schlichting}, Hilke E.},
        title = "{Sculpting the valley in the radius distribution of small exoplanets as a by-product of planet formation: the core-powered mass-loss mechanism}",
      journal = {\mnras},
         year = 2019,
        month = jul,
       volume = {487},
       number = {1},
        pages = {24-33},
          doi = {10.1093/mnras/stz1230},
archivePrefix = {arXiv},
       eprint = {1811.03202},
 primaryClass = {astro-ph.EP},
       adsurl = {https://ui.adsabs.harvard.edu/abs/2019MNRAS.487...24G}
}

@ARTICLE{Owen2017,
       author = {{Owen}, James E. and {Wu}, Yanqin},
        title = "{The Evaporation Valley in the Kepler Planets}",
      journal = {\apj},
         year = 2017,
        month = sep,
       volume = {847},
       number = {1},
          eid = {29},
        pages = {29},
          doi = {10.3847/1538-4357/aa890a},
archivePrefix = {arXiv},
       eprint = {1705.10810},
 primaryClass = {astro-ph.EP},
       adsurl = {https://ui.adsabs.harvard.edu/abs/2017ApJ...847...29O}
}

@ARTICLE{Mordasini2009,
       author = {{Mordasini}, C. and {Alibert}, Y. and {Benz}, W.},
        title = "{Extrasolar planet population synthesis. I. Method, formation tracks, and mass-distance distribution}",
      journal = {\aap},
         year = 2009,
        month = jul,
       volume = {501},
       number = {3},
        pages = {1139-1160},
          doi = {10.1051/0004-6361/200810301},
archivePrefix = {arXiv},
       eprint = {0904.2524},
 primaryClass = {astro-ph.EP},
       adsurl = {https://ui.adsabs.harvard.edu/abs/2009A&A...501.1139M}
}

@ARTICLE{Kite2019,
       author = {{Kite}, Edwin S. and {Fegley}, Jr., Bruce and {Schaefer}, Laura and {Ford}, Eric B.},
        title = "{Superabundance of Exoplanet Sub-Neptunes Explained by Fugacity Crisis}",
      journal = {\apjl},
         year = 2019,
        month = dec,
       volume = {887},
       number = {2},
          eid = {L33},
        pages = {L33},
          doi = {10.3847/2041-8213/ab59d9},
archivePrefix = {arXiv},
       eprint = {1912.02701},
 primaryClass = {astro-ph.EP},
       adsurl = {https://ui.adsabs.harvard.edu/abs/2019ApJ...887L..33K}
}

@ARTICLE{Nixon2024_unc,
       author = {{Nixon}, Matthew C. and {Welbanks}, Luis and {McGill}, Peter and {Kempton}, Eliza M. -R.},
        title = "{Methods for Incorporating Model Uncertainty into Exoplanet Atmospheric Analysis}",
      journal = {\apj},
         year = 2024,
        month = may,
       volume = {966},
       number = {2},
          eid = {156},
        pages = {156},
          doi = {10.3847/1538-4357/ad354e},
archivePrefix = {arXiv},
       eprint = {2310.03713},
 primaryClass = {astro-ph.EP},
       adsurl = {https://ui.adsabs.harvard.edu/abs/2024ApJ...966..156N}
}

@ARTICLE{Pinhas2018,
       author = {{Pinhas}, Arazi and {Rackham}, Benjamin V. and {Madhusudhan}, Nikku and {Apai}, D{\'a}niel},
        title = "{Retrieval of planetary and stellar properties in transmission spectroscopy with AURA}",
      journal = {\mnras},
         year = 2018,
        month = nov,
       volume = {480},
       number = {4},
        pages = {5314-5331},
          doi = {10.1093/mnras/sty2209},
archivePrefix = {arXiv},
       eprint = {1808.10017},
 primaryClass = {astro-ph.EP},
       adsurl = {https://ui.adsabs.harvard.edu/abs/2018MNRAS.480.5314P}
}

@ARTICLE{Husser2013,
       author = {{Husser}, T. -O. and {Wende-von Berg}, S. and {Dreizler}, S. and {Homeier}, D. and {Reiners}, A. and {Barman}, T. and {Hauschildt}, P.~H.},
        title = "{A new extensive library of PHOENIX stellar atmospheres and synthetic spectra}",
      journal = {\aap},
         year = 2013,
        month = may,
       volume = {553},
          eid = {A6},
        pages = {A6},
          doi = {10.1051/0004-6361/201219058},
archivePrefix = {arXiv},
       eprint = {1303.5632},
 primaryClass = {astro-ph.SR},
       adsurl = {https://ui.adsabs.harvard.edu/abs/2013A&A...553A...6H}
}

@ARTICLE{Nixon2022,
       author = {{Nixon}, Matthew C. and {Madhusudhan}, Nikku},
        title = "{Aura-3D: A Three-dimensional Atmospheric Retrieval Framework for Exoplanet Transmission Spectra}",
      journal = {\apj},
         year = 2022,
        month = aug,
       volume = {935},
       number = {2},
          eid = {73},
        pages = {73},
          doi = {10.3847/1538-4357/ac7c09},
archivePrefix = {arXiv},
       eprint = {2201.03532},
 primaryClass = {astro-ph.EP},
       adsurl = {https://ui.adsabs.harvard.edu/abs/2022ApJ...935...73N}
}

@ARTICLE{Schaefer2016,
       author = {{Schaefer}, Laura and {Wordsworth}, Robin D. and {Berta-Thompson}, Zachory and {Sasselov}, Dimitar},
        title = "{Predictions of the Atmospheric Composition of GJ 1132b}",
      journal = {\apj},
         year = 2016,
        month = oct,
       volume = {829},
       number = {2},
          eid = {63},
        pages = {63},
          doi = {10.3847/0004-637X/829/2/63},
archivePrefix = {arXiv},
       eprint = {1607.03906},
 primaryClass = {astro-ph.EP},
       adsurl = {https://ui.adsabs.harvard.edu/abs/2016ApJ...829...63S}
}

@ARTICLE{Shorttle2024,
       author = {{Shorttle}, Oliver and {Jordan}, Sean and {Nicholls}, Harrison and {Lichtenberg}, Tim and {Bower}, Dan J.},
        title = "{Distinguishing Oceans of Water from Magma on Mini-Neptune K2-18b}",
      journal = {\apjl},
         year = 2024,
        month = feb,
       volume = {962},
       number = {1},
          eid = {L8},
        pages = {L8},
          doi = {10.3847/2041-8213/ad206e},
archivePrefix = {arXiv},
       eprint = {2401.05864},
 primaryClass = {astro-ph.EP},
       adsurl = {https://ui.adsabs.harvard.edu/abs/2024ApJ...962L...8S}
}

@ARTICLE{Rigby2024,
       author = {{Rigby}, Frances E. and {Pica-Ciamarra}, Lorenzo and {Holmberg}, M{\r{a}}ns and {Madhusudhan}, Nikku and {Constantinou}, Savvas and {Schaefer}, Laura and {Deng}, Jie and {Lee}, Kanani K.~M. and {Moses}, Julianne I.},
        title = "{Toward a Self-consistent Evaluation of Gas Dwarf Scenarios for Temperate Sub-Neptunes}",
      journal = {\apj},
         year = 2024,
        month = nov,
       volume = {975},
       number = {1},
          eid = {101},
        pages = {101},
          doi = {10.3847/1538-4357/ad6c38},
archivePrefix = {arXiv},
       eprint = {2409.03683},
 primaryClass = {astro-ph.EP},
       adsurl = {https://ui.adsabs.harvard.edu/abs/2024ApJ...975..101R}
}

@ARTICLE{Seo2024,
       author = {{Seo}, Chanoul and {Ito}, Yuichi and {Fujii}, Yuka},
        title = "{Role of Magma Oceans in Controlling Carbon and Oxygen of Sub-Neptune Atmospheres}",
      journal = {\apj},
         year = 2024,
        month = nov,
       volume = {975},
       number = {1},
          eid = {14},
        pages = {14},
          doi = {10.3847/1538-4357/ad7461},
archivePrefix = {arXiv},
       eprint = {2408.17056},
 primaryClass = {astro-ph.EP},
       adsurl = {https://ui.adsabs.harvard.edu/abs/2024ApJ...975...14S}
}

@ARTICLE{Holmberg2024,
       author = {{Holmberg}, M{\r{a}}ns and {Madhusudhan}, Nikku},
        title = "{Possible Hycean conditions in the sub-Neptune TOI-270 d}",
      journal = {\aap},
         year = 2024,
        month = mar,
       volume = {683},
          eid = {L2},
        pages = {L2},
          doi = {10.1051/0004-6361/202348238},
archivePrefix = {arXiv},
       eprint = {2403.03244},
 primaryClass = {astro-ph.EP},
       adsurl = {https://ui.adsabs.harvard.edu/abs/2024A&A...683L...2H}
}

@ARTICLE{Davenport2025,
       author = {{Davenport}, Brian and {Kempton}, Eliza M. -R. and {Nixon}, Matthew C. and {Ih}, Jegug and {Deming}, Drake and {Fu}, Guangwei and {May}, E.~M. and {Bean}, Jacob L. and {Gao}, Peter and {Rogers}, Leslie and {Malik}, Matej},
        title = "{TOI-421 b: A Hot Sub-Neptune with a Haze-Free, Low Mean Molecular Weight Atmosphere}",
      journal = {arXiv e-prints},
         year = 2025,
        month = jan,
          eid = {arXiv:2501.01498},
        pages = {arXiv:2501.01498},
          doi = {10.48550/arXiv.2501.01498},
archivePrefix = {arXiv},
       eprint = {2501.01498},
 primaryClass = {astro-ph.EP},
       adsurl = {https://ui.adsabs.harvard.edu/abs/2025arXiv250101498D}
}

@ARTICLE{Piaulet2024,
       author = {{Piaulet-Ghorayeb}, Caroline and {Benneke}, Bj{\"o}rn and {Radica}, Michael and {Raul}, Eshan and {Coulombe}, Louis-Philippe and {Ahrer}, Eva-Maria and {Kubyshkina}, Daria and {Howard}, Ward S. and {Krissansen-Totton}, Joshua and {MacDonald}, Ryan J. and {Roy}, Pierre-Alexis and {Louca}, Amy and {Christie}, Duncan and {Fournier-Tondreau}, Marylou and {Allart}, Romain and {Miguel}, Yamila and {Schlichting}, Hilke E. and {Welbanks}, Luis and {Cadieux}, Charles and {Dorn}, Caroline and {Evans-Soma}, Thomas M. and {Fortney}, Jonathan J. and {Pierrehumbert}, Raymond and {Lafreni{\`e}re}, David and {Acu{\~n}a}, Lorena and {Komacek}, Thaddeus and {Innes}, Hamish and {Beatty}, Thomas G. and {Cloutier}, Ryan and {Doyon}, Ren{\'e} and {Gagnebin}, Anna and {Gapp}, Cyril and {Knutson}, Heather A.},
        title = "{JWST/NIRISS Reveals the Water-rich ``Steam World'' Atmosphere of GJ 9827 d}",
      journal = {\apjl},
         year = 2024,
        month = oct,
       volume = {974},
       number = {1},
          eid = {L10},
        pages = {L10},
          doi = {10.3847/2041-8213/ad6f00},
archivePrefix = {arXiv},
       eprint = {2410.03527},
 primaryClass = {astro-ph.EP},
       adsurl = {https://ui.adsabs.harvard.edu/abs/2024ApJ...974L..10P}
}

@ARTICLE{Schlawin2024,
       author = {{Schlawin}, Everett and {Ohno}, Kazumasa and {Bell}, Taylor J. and {Murphy}, Matthew M. and {Welbanks}, Luis and {Beatty}, Thomas G. and {Greene}, Thomas P. and {Fortney}, Jonathan J. and {Parmentier}, Vivien and {Edelman}, Isaac R. and {Gill}, Samuel and {Anderson}, David R. and {Wheatley}, Peter J. and {Henry}, Gregory W. and {Mehta}, Nishil and {Kreidberg}, Laura and {Rieke}, Marcia J.},
        title = "{Possible Carbon Dioxide above the Thick Aerosols of GJ 1214 b}",
      journal = {\apjl},
         year = 2024,
        month = oct,
       volume = {974},
       number = {2},
          eid = {L33},
        pages = {L33},
          doi = {10.3847/2041-8213/ad7fef},
archivePrefix = {arXiv},
       eprint = {2410.10183},
 primaryClass = {astro-ph.EP},
       adsurl = {https://ui.adsabs.harvard.edu/abs/2024ApJ...974L..33S}
}

@ARTICLE{Gunther2019,
       author = {{G{\"u}nther}, Maximilian N. and {Pozuelos}, Francisco J. and {Dittmann}, Jason A. and {Dragomir}, Diana and {Kane}, Stephen R. and {Daylan}, Tansu and {Feinstein}, Adina D. and {Huang}, Chelsea X. and {Morton}, Timothy D. and {Bonfanti}, Andrea and {Bouma}, L.~G. and {Burt}, Jennifer and {Collins}, Karen A. and {Lissauer}, Jack J. and {Matthews}, Elisabeth and {Montet}, Benjamin T. and {Vanderburg}, Andrew and {Wang}, Songhu and {Winters}, Jennifer G. and {Ricker}, George R. and {Vanderspek}, Roland K. and {Latham}, David W. and {Seager}, Sara and {Winn}, Joshua N. and {Jenkins}, Jon M. and {Armstrong}, James D. and {Barkaoui}, Khalid and {Batalha}, Natalie and {Bean}, Jacob L. and {Caldwell}, Douglas A. and {Ciardi}, David R. and {Collins}, Kevin I. and {Crossfield}, Ian and {Fausnaugh}, Michael and {Furesz}, Gabor and {Gan}, Tianjun and {Gillon}, Micha{\"e}l and {Guerrero}, Natalia and {Horne}, Keith and {Howell}, Steve B. and {Ireland}, Michael and {Isopi}, Giovanni and {Jehin}, Emmanu{\"e}l and {Kielkopf}, John F. and {Lepine}, Sebastien and {Mallia}, Franco and {Matson}, Rachel A. and {Myers}, Gordon and {Palle}, Enric and {Quinn}, Samuel N. and {Relles}, Howard M. and {Rojas-Ayala}, B{\'a}rbara and {Schlieder}, Joshua and {Sefako}, Ramotholo and {Shporer}, Avi and {Su{\'a}rez}, Juan C. and {Tan}, Thiam-Guan and {Ting}, Eric B. and {Twicken}, Joseph D. and {Waite}, Ian A.},
        title = "{A super-Earth and two sub-Neptunes transiting the nearby and quiet M dwarf TOI-270}",
      journal = {Nature Astronomy},
         year = 2019,
        month = jul,
       volume = {3},
        pages = {1099-1108},
          doi = {10.1038/s41550-019-0845-5},
archivePrefix = {arXiv},
       eprint = {1903.06107},
 primaryClass = {astro-ph.EP},
       adsurl = {https://ui.adsabs.harvard.edu/abs/2019NatAs...3.1099G}
}

@ARTICLE{MikalEvans2023,
       author = {{Mikal-Evans}, Thomas and {Madhusudhan}, Nikku and {Dittmann}, Jason and {G{\"u}nther}, Maximilian N. and {Welbanks}, Luis and {Van Eylen}, Vincent and {Crossfield}, Ian J.~M. and {Daylan}, Tansu and {Kreidberg}, Laura},
        title = "{Hubble Space Telescope Transmission Spectroscopy for the Temperate Sub-Neptune TOI-270 d: A Possible Hydrogen-rich Atmosphere Containing Water Vapor}",
      journal = {\aj},
         year = 2023,
        month = mar,
       volume = {165},
       number = {3},
          eid = {84},
        pages = {84},
          doi = {10.3847/1538-3881/aca90b},
archivePrefix = {arXiv},
       eprint = {2211.15576},
 primaryClass = {astro-ph.EP},
       adsurl = {https://ui.adsabs.harvard.edu/abs/2023AJ....165...84M}
}

@ARTICLE{Barton2013,
       author = {{Barton}, Emma J. and {Yurchenko}, Sergei N. and {Tennyson}, Jonathan},
        title = "{ExoMol line lists - II. The ro-vibrational spectrum of SiO}",
      journal = {\mnras},
         year = 2013,
        month = sep,
       volume = {434},
       number = {2},
        pages = {1469-1475},
          doi = {10.1093/mnras/stt1105},
archivePrefix = {arXiv},
       eprint = {1307.2300},
 primaryClass = {astro-ph.SR},
       adsurl = {https://ui.adsabs.harvard.edu/abs/2013MNRAS.434.1469B}
}

@ARTICLE{Owens2017,
       author = {{Owens}, A. and {Yachmenev}, A. and {Thiel}, W. and {Tennyson}, J. and {Yurchenko}, S.~N.},
        title = "{ExoMol line lists - XXII. The rotation-vibration spectrum of silane up to 1200 K}",
      journal = {\mnras},
         year = 2017,
        month = nov,
       volume = {471},
       number = {4},
        pages = {5025-5032},
          doi = {10.1093/mnras/stx1952},
archivePrefix = {arXiv},
       eprint = {1712.09354},
 primaryClass = {physics.chem-ph},
       adsurl = {https://ui.adsabs.harvard.edu/abs/2017MNRAS.471.5025O}
}

@ARTICLE{SousaSilva2015,
       author = {{Sousa-Silva}, Clara and {Al-Refaie}, Ahmed F. and {Tennyson}, Jonathan and {Yurchenko}, Sergei N.},
        title = "{ExoMol line lists - VII. The rotation-vibration spectrum of phosphine up to 1500 K}",
      journal = {\mnras},
         year = 2015,
        month = jan,
       volume = {446},
       number = {3},
        pages = {2337-2347},
          doi = {10.1093/mnras/stu2246},
archivePrefix = {arXiv},
       eprint = {1410.2917},
 primaryClass = {astro-ph.EP},
       adsurl = {https://ui.adsabs.harvard.edu/abs/2015MNRAS.446.2337S}
}

@ARTICLE{HITRAN2020,
       author = {{Gordon}, I.~E. and {Rothman}, L.~S. and {Hargreaves}, R.~J. and {Hashemi}, R. and {Karlovets}, E.~V. and {Skinner}, F.~M. and {Conway}, E.~K. and {Hill}, C. and {Kochanov}, R.~V. and {Tan}, Y. and {Wcis{\l}o}, P. and {Finenko}, A.~A. and {Nelson}, K. and {Bernath}, P.~F. and {Birk}, M. and {Boudon}, V. and {Campargue}, A. and {Chance}, K.~V. and {Coustenis}, A. and {Drouin}, B.~J. and {Flaud}, J. -M. and {Gamache}, R.~R. and {Hodges}, J.~T. and {Jacquemart}, D. and {Mlawer}, E.~J. and {Nikitin}, A.~V. and {Perevalov}, V.~I. and {Rotger}, M. and {Tennyson}, J. and {Toon}, G.~C. and {Tran}, H. and {Tyuterev}, V.~G. and {Adkins}, E.~M. and {Baker}, A. and {Barbe}, A. and {Can{\`e}}, E. and {Cs{\'a}sz{\'a}r}, A.~G. and {Dudaryonok}, A. and {Egorov}, O. and {Fleisher}, A.~J. and {Fleurbaey}, H. and {Foltynowicz}, A. and {Furtenbacher}, T. and {Harrison}, J.~J. and {Hartmann}, J. -M. and {Horneman}, V. -M. and {Huang}, X. and {Karman}, T. and {Karns}, J. and {Kassi}, S. and {Kleiner}, I. and {Kofman}, V. and {Kwabia-Tchana}, F. and {Lavrentieva}, N.~N. and {Lee}, T.~J. and {Long}, D.~A. and {Lukashevskaya}, A.~A. and {Lyulin}, O.~M. and {Makhnev}, V. Yu. and {Matt}, W. and {Massie}, S.~T. and {Melosso}, M. and {Mikhailenko}, S.~N. and {Mondelain}, D. and {M{\"u}ller}, H.~S.~P. and {Naumenko}, O.~V. and {Perrin}, A. and {Polyansky}, O.~L. and {Raddaoui}, E. and {Raston}, P.~L. and {Reed}, Z.~D. and {Rey}, M. and {Richard}, C. and {T{\'o}bi{\'a}s}, R. and {Sadiek}, I. and {Schwenke}, D.~W. and {Starikova}, E. and {Sung}, K. and {Tamassia}, F. and {Tashkun}, S.~A. and {Vander Auwera}, J. and {Vasilenko}, I.~A. and {Vigasin}, A.~A. and {Villanueva}, G.~L. and {Vispoel}, B. and {Wagner}, G. and {Yachmenev}, A. and {Yurchenko}, S.~N.},
        title = "{The HITRAN2020 molecular spectroscopic database}",
      journal = {\jqsrt},
         year = 2022,
        month = jan,
       volume = {277},
          eid = {107949},
        pages = {107949},
          doi = {10.1016/j.jqsrt.2021.107949},
       adsurl = {https://ui.adsabs.harvard.edu/abs/2022JQSRT.27707949G}
}

@ARTICLE{Coles2019,
       author = {{Coles}, Phillip A. and {Yurchenko}, Sergei N. and {Tennyson}, Jonathan},
        title = "{ExoMol molecular line lists - XXXV. A rotation-vibration line list for hot ammonia}",
      journal = {\mnras},
         year = 2019,
        month = dec,
       volume = {490},
       number = {4},
        pages = {4638-4647},
          doi = {10.1093/mnras/stz2778},
archivePrefix = {arXiv},
       eprint = {1911.10369},
 primaryClass = {astro-ph.SR},
       adsurl = {https://ui.adsabs.harvard.edu/abs/2019MNRAS.490.4638C}
}

@ARTICLE{Chubb2018,
       author = {{Chubb}, Katy L. and {Naumenko}, Olga and {Keely}, Stefan and {Bartolotto}, Sebestiano and {Macdonald}, Skye and {Mukhtar}, Mahmoud and {Grachov}, Andrey and {White}, Joe and {Coleman}, Eden and {Liu}, Anwen and {Fazliev}, Alexander Z. and {Polovtseva}, Elena R. and {Horneman}, Veli-Matti and {Campargue}, Alain and {Furtenbacher}, Tibor and {Cs{\'a}sz{\'a}r}, Attila G. and {Yurchenko}, Sergei N. and {Tennyson}, Jonathan},
        title = "{MARVEL analysis of the measured high-resolution rovibrational spectra of H$_{2}$$^{32}$S}",
      journal = {\jqsrt},
         year = 2018,
        month = oct,
       volume = {218},
        pages = {178-186},
          doi = {10.1016/j.jqsrt.2018.07.012},
archivePrefix = {arXiv},
       eprint = {1812.10503},
 primaryClass = {astro-ph.EP},
       adsurl = {https://ui.adsabs.harvard.edu/abs/2018JQSRT.218..178C}
}

@ARTICLE{Azzam2016,
       author = {{Azzam}, Ala'a. A.~A. and {Tennyson}, Jonathan and {Yurchenko}, Sergei N. and {Naumenko}, Olga V.},
        title = "{ExoMol molecular line lists - XVI. The rotation-vibration spectrum of hot H$_{2}$S}",
      journal = {\mnras},
         year = 2016,
        month = aug,
       volume = {460},
       number = {4},
        pages = {4063-4074},
          doi = {10.1093/mnras/stw1133},
archivePrefix = {arXiv},
       eprint = {1607.00499},
 primaryClass = {astro-ph.EP},
       adsurl = {https://ui.adsabs.harvard.edu/abs/2016MNRAS.460.4063A}
}

@ARTICLE{Polyansky2018,
       author = {{Polyansky}, Oleg L. and {Kyuberis}, Aleksandra A. and {Zobov}, Nikolai F. and {Tennyson}, Jonathan and {Yurchenko}, Sergei N. and {Lodi}, Lorenzo},
        title = "{ExoMol molecular line lists XXX: a complete high-accuracy line list for water}",
      journal = {\mnras},
         year = 2018,
        month = oct,
       volume = {480},
       number = {2},
        pages = {2597-2608},
          doi = {10.1093/mnras/sty1877},
archivePrefix = {arXiv},
       eprint = {1807.04529},
 primaryClass = {astro-ph.EP},
       adsurl = {https://ui.adsabs.harvard.edu/abs/2018MNRAS.480.2597P}
}

@BOOK{Kurucz1995,
       author = {{Kurucz}, Robert L. and {Bell}, Barbara},
        title = "{Atomic line list}",
    publisher = {Smithsonian Astrophysical Observartory},
         year = 1995,
       adsurl = {https://ui.adsabs.harvard.edu/abs/1995all..book.....K}
}

@ARTICLE{Hargreaves2020,
       author = {{Hargreaves}, Robert J. and {Gordon}, Iouli E. and {Rey}, Michael and {Nikitin}, Andrei V. and {Tyuterev}, Vladimir G. and {Kochanov}, Roman V. and {Rothman}, Laurence S.},
        title = "{An Accurate, Extensive, and Practical Line List of Methane for the HITEMP Database}",
      journal = {\apjs},
         year = 2020,
        month = apr,
       volume = {247},
       number = {2},
          eid = {55},
        pages = {55},
          doi = {10.3847/1538-4365/ab7a1a},
archivePrefix = {arXiv},
       eprint = {2001.05037},
 primaryClass = {astro-ph.EP},
       adsurl = {https://ui.adsabs.harvard.edu/abs/2020ApJS..247...55H}
}

@ARTICLE{Chubb2020,
       author = {{Chubb}, Katy L. and {Tennyson}, Jonathan and {Yurchenko}, Sergei N.},
        title = "{ExoMol molecular line lists - XXXVII. Spectra of acetylene}",
      journal = {\mnras},
         year = 2020,
        month = apr,
       volume = {493},
       number = {2},
        pages = {1531-1545},
          doi = {10.1093/mnras/staa229},
archivePrefix = {arXiv},
       eprint = {2001.04550},
 primaryClass = {astro-ph.SR},
       adsurl = {https://ui.adsabs.harvard.edu/abs/2020MNRAS.493.1531C}
}

@ARTICLE{VanEylen2021,
       author = {{Van Eylen}, V. and {Astudillo-Defru}, N. and {Bonfils}, X. and {Livingston}, J. and {Hirano}, T. and {Luque}, R. and {Lam}, K.~W.~F. and {Justesen}, A.~B. and {Winn}, J.~N. and {Gandolfi}, D. and {Nowak}, G. and {Palle}, E. and {Albrecht}, S. and {Dai}, F. and {Campos Estrada}, B. and {Owen}, J.~E. and {Foreman-Mackey}, D. and {Fridlund}, M. and {Korth}, J. and {Mathur}, S. and {Forveille}, T. and {Mikal-Evans}, T. and {Osborne}, H.~L.~M. and {Ho}, C.~S.~K. and {Almenara}, J.~M. and {Artigau}, E. and {Barrag{\'a}n}, O. and {Barros}, S.~C.~C. and {Bouchy}, F. and {Cabrera}, J. and {Caldwell}, D.~A. and {Charbonneau}, D. and {Chaturvedi}, P. and {Cochran}, W.~D. and {Csizmadia}, S. and {Damasso}, M. and {Delfosse}, X. and {De Medeiros}, J.~R. and {D{\'\i}az}, R.~F. and {Doyon}, R. and {Esposito}, M. and {F{\H{u}}r{\'e}sz}, G. and {Figueira}, P. and {Georgieva}, I. and {Goffo}, E. and {Grziwa}, S. and {Guenther}, E. and {Hatzes}, A.~P. and {Jenkins}, J.~M. and {Kabath}, P. and {Knudstrup}, E. and {Latham}, D.~W. and {Lavie}, B. and {Lovis}, C. and {Mennickent}, R.~E. and {Mullally}, S.~E. and {Murgas}, F. and {Narita}, N. and {Pepe}, F.~A. and {Persson}, C.~M. and {Redfield}, S. and {Ricker}, G.~R. and {Santos}, N.~C. and {Seager}, S. and {Serrano}, L.~M. and {Smith}, A.~M.~S. and {Su{\'a}rez Mascare{\~n}o}, A. and {Subjak}, J. and {Twicken}, J.~D. and {Udry}, S. and {Vanderspek}, R. and {Zapatero Osorio}, M.~R.},
        title = "{Masses and compositions of three small planets orbiting the nearby M dwarf L231-32 (TOI-270) and the M dwarf radius valley}",
      journal = {\mnras},
         year = 2021,
        month = oct,
       volume = {507},
       number = {2},
        pages = {2154-2173},
          doi = {10.1093/mnras/stab2143},
archivePrefix = {arXiv},
       eprint = {2101.01593},
 primaryClass = {astro-ph.EP},
       adsurl = {https://ui.adsabs.harvard.edu/abs/2021MNRAS.507.2154V}
}

@ARTICLE{Chen2016,
       author = {{Chen}, Howard and {Rogers}, Leslie A.},
        title = "{Evolutionary Analysis of Gaseous Sub-Neptune-mass Planets with MESA}",
      journal = {\apj},
         year = 2016,
        month = nov,
       volume = {831},
       number = {2},
          eid = {180},
        pages = {180},
          doi = {10.3847/0004-637X/831/2/180},
archivePrefix = {arXiv},
       eprint = {1603.06596},
 primaryClass = {astro-ph.EP},
       adsurl = {https://ui.adsabs.harvard.edu/abs/2016ApJ...831..180C}
}

@ARTICLE{Kitzmann2024,
       author = {{Kitzmann}, Daniel and {Stock}, Joachim W. and {Patzer}, A. Beate C.},
        title = "{FASTCHEM COND: equilibrium chemistry with condensation and rainout for cool planetary and stellar environments}",
      journal = {\mnras},
         year = 2024,
        month = jan,
       volume = {527},
       number = {3},
        pages = {7263-7283},
          doi = {10.1093/mnras/stad3515},
archivePrefix = {arXiv},
       eprint = {2309.02337},
 primaryClass = {astro-ph.EP},
       adsurl = {https://ui.adsabs.harvard.edu/abs/2024MNRAS.527.7263K}
}

@ARTICLE{Wogan2023,
       author = {{Wogan}, Nicholas F. and {Catling}, David C. and {Zahnle}, Kevin J. and {Lupu}, Roxana},
        title = "{Origin-of-life Molecules in the Atmosphere after Big Impacts on the Early Earth}",
      journal = {\psj},
         year = 2023,
        month = sep,
       volume = {4},
       number = {9},
          eid = {169},
        pages = {169},
          doi = {10.3847/PSJ/aced83},
archivePrefix = {arXiv},
       eprint = {2307.09761},
 primaryClass = {astro-ph.EP},
       adsurl = {https://ui.adsabs.harvard.edu/abs/2023PSJ.....4..169W}
}

@ARTICLE{Malik2017,
       author = {{Malik}, Matej and {Grosheintz}, Luc and {Mendon{\c{c}}a}, Jo{\~a}o M. and {Grimm}, Simon L. and {Lavie}, Baptiste and {Kitzmann}, Daniel and {Tsai}, Shang-Min and {Burrows}, Adam and {Kreidberg}, Laura and {Bedell}, Megan and {Bean}, Jacob L. and {Stevenson}, Kevin B. and {Heng}, Kevin},
        title = "{HELIOS: An Open-source, GPU-accelerated Radiative Transfer Code for Self-consistent Exoplanetary Atmospheres}",
      journal = {\aj},
         year = 2017,
        month = feb,
       volume = {153},
       number = {2},
          eid = {56},
        pages = {56},
          doi = {10.3847/1538-3881/153/2/56},
archivePrefix = {arXiv},
       eprint = {1606.05474},
 primaryClass = {astro-ph.EP},
       adsurl = {https://ui.adsabs.harvard.edu/abs/2017AJ....153...56M}
}

@ARTICLE{Gupta2024,
       author = {{Gupta}, Akash and {Stixrude}, Lars and {Schlichting}, Hilke E.},
        title = "{The miscibility of hydrogen and water in planetary atmospheres and interiors}",
      journal = {arXiv e-prints},
         year = 2024,
        month = jul,
          eid = {arXiv:2407.04685},
        pages = {arXiv:2407.04685},
          doi = {10.48550/arXiv.2407.04685},
archivePrefix = {arXiv},
       eprint = {2407.04685},
 primaryClass = {astro-ph.EP},
       adsurl = {https://ui.adsabs.harvard.edu/abs/2024arXiv240704685G}
}

@ARTICLE{Benneke2024,
       author = {{Benneke}, Bj{\"o}rn and {Roy}, Pierre-Alexis and {Coulombe}, Louis-Philippe and {Radica}, Michael and {Piaulet}, Caroline and {Ahrer}, Eva-Maria and {Pierrehumbert}, Raymond and {Krissansen-Totton}, Joshua and {Schlichting}, Hilke E. and {Hu}, Renyu and {Yang}, Jeehyun and {Christie}, Duncan and {Thorngren}, Daniel and {Young}, Edward D. and {Pelletier}, Stefan and {Knutson}, Heather A. and {Miguel}, Yamila and {Evans-Soma}, Thomas M. and {Dorn}, Caroline and {Gagnebin}, Anna and {Fortney}, Jonathan J. and {Komacek}, Thaddeus and {MacDonald}, Ryan and {Raul}, Eshan and {Cloutier}, Ryan and {Acuna}, Lorena and {Lafreni{\`e}re}, David and {Cadieux}, Charles and {Doyon}, Ren{\'e} and {Welbanks}, Luis and {Allart}, Romain},
        title = "{JWST Reveals CH$_4$, CO$_2$, and H$_2$O in a Metal-rich Miscible Atmosphere on a Two-Earth-Radius Exoplanet}",
      journal = {arXiv e-prints},
         year = 2024,
        month = mar,
          eid = {arXiv:2403.03325},
        pages = {arXiv:2403.03325},
          doi = {10.48550/arXiv.2403.03325},
archivePrefix = {arXiv},
       eprint = {2403.03325},
 primaryClass = {astro-ph.EP},
       adsurl = {https://ui.adsabs.harvard.edu/abs/2024arXiv240303325B}
}

@ARTICLE{Madhusudhan2023,
       author = {{Madhusudhan}, Nikku and {Sarkar}, Subhajit and {Constantinou}, Savvas and {Holmberg}, M{\r{a}}ns and {Piette}, Anjali A.~A. and {Moses}, Julianne I.},
        title = "{Carbon-bearing Molecules in a Possible Hycean Atmosphere}",
      journal = {\apjl},
         year = 2023,
        month = oct,
       volume = {956},
       number = {1},
          eid = {L13},
        pages = {L13},
          doi = {10.3847/2041-8213/acf577},
archivePrefix = {arXiv},
       eprint = {2309.05566},
 primaryClass = {astro-ph.EP},
       adsurl = {https://ui.adsabs.harvard.edu/abs/2023ApJ...956L..13M}
}

@ARTICLE{Schlichting2022,
       author = {{Schlichting}, Hilke E. and {Young}, Edward D.},
        title = "{Chemical Equilibrium between Cores, Mantles, and Atmospheres of Super-Earths and Sub-Neptunes and Implications for Their Compositions, Interiors, and Evolution}",
      journal = {\psj},
         year = 2022,
        month = may,
       volume = {3},
       number = {5},
          eid = {127},
        pages = {127},
          doi = {10.3847/PSJ/ac68e6},
archivePrefix = {arXiv},
       eprint = {2107.10405},
 primaryClass = {astro-ph.EP},
       adsurl = {https://ui.adsabs.harvard.edu/abs/2022PSJ.....3..127S}
}

@ARTICLE{Ginzburg2016,
       author = {{Ginzburg}, Sivan and {Schlichting}, Hilke E. and {Sari}, Re'em},
        title = "{Super-Earth Atmospheres: Self-consistent Gas Accretion and Retention}",
      journal = {\apj},
         year = 2016,
        month = jul,
       volume = {825},
       number = {1},
          eid = {29},
        pages = {29},
          doi = {10.3847/0004-637X/825/1/29},
archivePrefix = {arXiv},
       eprint = {1512.07925},
 primaryClass = {astro-ph.EP},
       adsurl = {https://ui.adsabs.harvard.edu/abs/2016ApJ...825...29G}
}

@Article{Hunter2007,
  Author    = {Hunter, J. D.},
  Title     = {Matplotlib: A 2D graphics environment},
  Journal   = {Computing in Science \& Engineering},
  Volume    = {9},
  Number    = {3},
  Pages     = {90--95},
  publisher = {IEEE COMPUTER SOC},
  doi       = {10.1109/MCSE.2007.55},
  year      = 2007
}

@ARTICLE{Virtanen2020,
  author  = {Virtanen, Pauli and Gommers, Ralf and Oliphant, Travis E. and
            Haberland, Matt and Reddy, Tyler and Cournapeau, David and
            Burovski, Evgeni and Peterson, Pearu and Weckesser, Warren and
            Bright, Jonathan and {van der Walt}, St{\'e}fan J. and
            Brett, Matthew and Wilson, Joshua and Millman, K. Jarrod and
            Mayorov, Nikolay and Nelson, Andrew R. J. and Jones, Eric and
            Kern, Robert and Larson, Eric and Carey, C J and
            Polat, {\.I}lhan and Feng, Yu and Moore, Eric W. and
            {VanderPlas}, Jake and Laxalde, Denis and Perktold, Josef and
            Cimrman, Robert and Henriksen, Ian and Quintero, E. A. and
            Harris, Charles R. and Archibald, Anne M. and
            Ribeiro, Ant{\^o}nio H. and Pedregosa, Fabian and
            {van Mulbregt}, Paul and {SciPy 1.0 Contributors}},
  title   = {{{SciPy} 1.0: Fundamental Algorithms for Scientific
            Computing in Python}},
  journal = {Nature Methods},
  year    = {2020},
  volume  = {17},
  pages   = {261--272},
  adsurl  = {https://rdcu.be/b08Wh},
  doi     = {10.1038/s41592-019-0686-2},
}

@Article{Harris2020,
 title         = {Array programming with {NumPy}},
 author        = {Charles R. Harris and K. Jarrod Millman and St{\'{e}}fan J.
                 van der Walt and Ralf Gommers and Pauli Virtanen and David
                 Cournapeau and Eric Wieser and Julian Taylor and Sebastian
                 Berg and Nathaniel J. Smith and Robert Kern and Matti Picus
                 and Stephan Hoyer and Marten H. van Kerkwijk and Matthew
                 Brett and Allan Haldane and Jaime Fern{\'{a}}ndez del
                 R{\'{i}}o and Mark Wiebe and Pearu Peterson and Pierre
                 G{\'{e}}rard-Marchant and Kevin Sheppard and Tyler Reddy and
                 Warren Weckesser and Hameer Abbasi and Christoph Gohlke and
                 Travis E. Oliphant},
 year          = {2020},
 month         = sep,
 journal       = {Nature},
 volume        = {585},
 number        = {7825},
 pages         = {357--362},
 doi           = {10.1038/s41586-020-2649-2},
 publisher     = {Springer Science and Business Media {LLC}},
 url           = {https://doi.org/10.1038/s41586-020-2649-2}
}

@ARTICLE{Luque2022,
       author = {{Luque}, Rafael and {Pall{\'e}}, Enric},
        title = "{Density, not radius, separates rocky and water-rich small planets orbiting M dwarf stars}",
      journal = {Science},
         year = 2022,
        month = sep,
       volume = {377},
       number = {6611},
        pages = {1211-1214},
          doi = {10.1126/science.abl7164},
archivePrefix = {arXiv},
       eprint = {2209.03871},
 primaryClass = {astro-ph.EP},
       adsurl = {https://ui.adsabs.harvard.edu/abs/2022Sci...377.1211L}
}

@ARTICLE{Gao2023,
       author = {{Gao}, Peter and {Piette}, Anjali A.~A. and {Steinrueck}, Maria E. and {Nixon}, Matthew C. and {Zhang}, Michael and {Kempton}, Eliza M. -R. and {Bean}, Jacob L. and {Rauscher}, Emily and {Parmentier}, Vivien and {Batalha}, Natasha E. and {Savel}, Arjun B. and {Arnold}, Kenneth E. and {Roman}, Michael T. and {Malsky}, Isaac and {Taylor}, Jake},
        title = "{The Hazy and Metal-rich Atmosphere of GJ 1214 b Constrained by Near- and Mid-infrared Transmission Spectroscopy}",
      journal = {\apj},
         year = 2023,
        month = jul,
       volume = {951},
       number = {2},
          eid = {96},
        pages = {96},
          doi = {10.3847/1538-4357/acd16f},
archivePrefix = {arXiv},
       eprint = {2305.05697},
 primaryClass = {astro-ph.EP},
       adsurl = {https://ui.adsabs.harvard.edu/abs/2023ApJ...951...96G}
}

@ARTICLE{Kempton2023_ww,
       author = {{Kempton}, Eliza M. -R. and {Lessard}, Madeline and {Malik}, Matej and {Rogers}, Leslie A. and {Futrowsky}, Kate E. and {Ih}, Jegug and {Marounina}, Nadejda and {Romero-Mirza}, Carlos E.},
        title = "{Where are the Water Worlds?: Self-consistent Models of Water-rich Exoplanet Atmospheres}",
      journal = {\apj},
         year = 2023,
        month = aug,
       volume = {953},
       number = {1},
          eid = {57},
        pages = {57},
          doi = {10.3847/1538-4357/ace10d},
archivePrefix = {arXiv},
       eprint = {2307.06508},
 primaryClass = {astro-ph.EP},
       adsurl = {https://ui.adsabs.harvard.edu/abs/2023ApJ...953...57K}
}

@ARTICLE{Kempton2023_gj,
       author = {{Kempton}, Eliza M. -R. and {Zhang}, Michael and {Bean}, Jacob L. and {Steinrueck}, Maria E. and {Piette}, Anjali A.~A. and {Parmentier}, Vivien and {Malsky}, Isaac and {Roman}, Michael T. and {Rauscher}, Emily and {Gao}, Peter and {Bell}, Taylor J. and {Xue}, Qiao and {Taylor}, Jake and {Savel}, Arjun B. and {Arnold}, Kenneth E. and {Nixon}, Matthew C. and {Stevenson}, Kevin B. and {Mansfield}, Megan and {Kendrew}, Sarah and {Zieba}, Sebastian and {Ducrot}, Elsa and {Dyrek}, Achr{\`e}ne and {Lagage}, Pierre-Olivier and {Stassun}, Keivan G. and {Henry}, Gregory W. and {Barman}, Travis and {Lupu}, Roxana and {Malik}, Matej and {Kataria}, Tiffany and {Ih}, Jegug and {Fu}, Guangwei and {Welbanks}, Luis and {McGill}, Peter},
        title = "{A reflective, metal-rich atmosphere for GJ 1214b from its JWST phase curve}",
      journal = {\nat},
         year = 2023,
        month = aug,
       volume = {620},
       number = {7972},
        pages = {67-71},
          doi = {10.1038/s41586-023-06159-5},
archivePrefix = {arXiv},
       eprint = {2305.06240},
 primaryClass = {astro-ph.EP},
       adsurl = {https://ui.adsabs.harvard.edu/abs/2023Natur.620...67K}
}

@ARTICLE{Asplund2009,
       author = {{Asplund}, Martin and {Grevesse}, Nicolas and {Sauval}, A. Jacques and
         {Scott}, Pat},
        title = "{The Chemical Composition of the Sun}",
      journal = {\araa},
         year = 2009,
        month = sep,
       volume = {47},
       number = {1},
        pages = {481-522},
          doi = {10.1146/annurev.astro.46.060407.145222},
archivePrefix = {arXiv},
       eprint = {0909.0948},
 primaryClass = {astro-ph.SR},
       adsurl = {https://ui.adsabs.harvard.edu/abs/2009ARA&A..47..481A}
}

@ARTICLE{Fulton2017,
       author = {{Fulton}, Benjamin J. and {Petigura}, Erik A. and {Howard}, Andrew W. and
         {Isaacson}, Howard and {Marcy}, Geoffrey W. and {Cargile}, Phillip A. and
         {Hebb}, Leslie and {Weiss}, Lauren M. and {Johnson}, John Asher and
         {Morton}, Timothy D. and {Sinukoff}, Evan and {Crossfield}, Ian J.~M. and
         {Hirsch}, Lea A.},
        title = "{The California-Kepler Survey. III. A Gap in the Radius Distribution of Small Planets}",
      journal = {\aj},
         year = 2017,
        month = sep,
       volume = {154},
       number = {3},
          eid = {109},
        pages = {109},
          doi = {10.3847/1538-3881/aa80eb},
archivePrefix = {arXiv},
       eprint = {1703.10375},
 primaryClass = {astro-ph.EP},
       adsurl = {https://ui.adsabs.harvard.edu/abs/2017AJ....154..109F}
}

@ARTICLE{Misener2023,
       author = {{Misener}, William and {Schlichting}, Hilke E. and {Young}, Edward D.},
        title = "{Atmospheres as windows into sub-Neptune interiors: coupled chemistry and structure of hydrogen-silane-water envelopes}",
      journal = {\mnras},
         year = 2023,
        month = sep,
       volume = {524},
       number = {1},
        pages = {981-992},
          doi = {10.1093/mnras/stad1910},
archivePrefix = {arXiv},
       eprint = {2303.09653},
 primaryClass = {astro-ph.EP},
       adsurl = {https://ui.adsabs.harvard.edu/abs/2023MNRAS.524..981M}
}

@ARTICLE{Welbanks2024,
       author = {{Welbanks}, Luis and {Bell}, Taylor J. and {Beatty}, Thomas G. and {Line}, Michael R. and {Ohno}, Kazumasa and {Fortney}, Jonathan J. and {Schlawin}, Everett and {Greene}, Thomas P. and {Rauscher}, Emily and {McGill}, Peter and {Murphy}, Matthew and {Parmentier}, Vivien and {Tang}, Yao and {Edelman}, Isaac and {Mukherjee}, Sagnick and {Wiser}, Lindsey S. and {Lagage}, Pierre-Olivier and {Dyrek}, Achr{\`e}ne and {Arnold}, Kenneth E.},
        title = "{A high internal heat flux and large core in a warm Neptune exoplanet}",
      journal = {\nat},
         year = 2024,
        month = jun,
       volume = {630},
       number = {8018},
        pages = {836-840},
          doi = {10.1038/s41586-024-07514-w},
archivePrefix = {arXiv},
       eprint = {2405.11018},
 primaryClass = {astro-ph.EP},
       adsurl = {https://ui.adsabs.harvard.edu/abs/2024Natur.630..836W}
}

@ARTICLE{Nixon2024_gj,
       author = {{Nixon}, Matthew C. and {Piette}, Anjali A.~A. and {Kempton}, Eliza M. -R. and {Gao}, Peter and {Bean}, Jacob L. and {Steinrueck}, Maria E. and {Mahajan}, Alexandra S. and {Eastman}, Jason D. and {Zhang}, Michael and {Rogers}, Leslie A.},
        title = "{New Insights into the Internal Structure of GJ 1214 b Informed by JWST}",
      journal = {\apjl},
         year = 2024,
        month = aug,
       volume = {970},
       number = {2},
          eid = {L28},
        pages = {L28},
          doi = {10.3847/2041-8213/ad615b},
archivePrefix = {arXiv},
       eprint = {2407.12079},
 primaryClass = {astro-ph.EP},
       adsurl = {https://ui.adsabs.harvard.edu/abs/2024ApJ...970L..28N}
}

@ARTICLE{Journaux2020,
       author = {{Journaux}, Baptiste and {Kalousov{\'a}}, Kl{\'a}ra and
         {Sotin}, Christophe and {Tobie}, Gabriel and {Vance}, Steve and
         {Saur}, Joachim and {Bollengier}, Olivier and {Noack}, Lena and
         {R{\"u}ckriemen-Bez}, Tina and {Van Hoolst}, Tim and
         {Soderlund}, Krista M. and {Brown}, J. Michael},
        title = "{Large Ocean Worlds with High-Pressure Ices}",
      journal = {\ssr},
         year = 2020,
        month = jan,
       volume = {216},
       number = {1},
          eid = {7},
        pages = {7},
          doi = {10.1007/s11214-019-0633-7},
       adsurl = {https://ui.adsabs.harvard.edu/abs/2020SSRv..216....7J}
}

@ARTICLE{Madhu2020,
       author = {{Madhusudhan}, Nikku and {Nixon}, Matthew C. and {Welbanks}, Luis and
         {Piette}, Anjali A.~A. and {Booth}, Richard A.},
        title = "{The Interior and Atmosphere of the Habitable-zone Exoplanet K2-18b}",
      journal = {\apjl},
         year = 2020,
        month = mar,
       volume = {891},
       number = {1},
          eid = {L7},
        pages = {L7},
          doi = {10.3847/2041-8213/ab7229},
       adsurl = {https://ui.adsabs.harvard.edu/abs/2020ApJ...891L...7M}
}

@ARTICLE{Rogers2015,
       author = {{Rogers}, Leslie A.},
        title = "{Most 1.6 Earth-radius Planets are Not Rocky}",
      journal = {\apj},
         year = 2015,
        month = mar,
       volume = {801},
       number = {1},
          eid = {41},
        pages = {41},
          doi = {10.1088/0004-637X/801/1/41},
archivePrefix = {arXiv},
       eprint = {1407.4457},
 primaryClass = {astro-ph.EP},
       adsurl = {https://ui.adsabs.harvard.edu/abs/2015ApJ...801...41R}
}
\bibliographystyle{aasjournalv7}

\end{document}